\documentclass[12pt,a4paper]{article}
\usepackage{graphicx}
\usepackage{comment}   
\usepackage[T1]{fontenc}
\usepackage[utf8]{inputenc}
\usepackage{textcomp}
\usepackage[sc,osf]{mathpazo}
\usepackage{a4wide}  
\usepackage{latexsym,amsthm,amsfonts,amsmath,mathrsfs,amssymb}
\usepackage{dsfont}
\usepackage{accents}
\usepackage[nosort]{cite}
\usepackage{booktabs} 
\usepackage[unicode,implicit]{hyperref}
\hypersetup{%
  pdftitle    = {The first law of heterotic stringy black hole mechanics at zeroth order  in $\alpha'$}
  pdfkeywords = {duality, T duality, strings, heterotic superstrings, effective action, gauge symmetry, entropy,  black holes, Noether, Wald, compactification, momentum maps, thermodynamics}
  pdfauthor   = {Zachary Elgood, Dimitrios Mitsios, Tom\'as Ort\'{\i}n and David Pere\~n\'{\i}guez},
  plainpages  = true,
  colorlinks  = true,
  citecolor   = blue,
  urlcolor    = red,
  linkcolor   = black
}
\newcommand{\hepth}[1]{{\tt
\href{http://www.arXiv.org/abs/hep-th/#1}{hep-th/#1}}}
\newcommand{\grqc}[1]{{\tt
\href{http://www.arXiv.org/abs/gr-qc/#1}{gr-qc/#1}}}

\newcommand{\arxiv}[1]{{\tt arXiv:\href{http://www.arXiv.org/abs/#1}{#1}}}
\newcommand{\doi}[1]{{\tt DOI:\href{http://dx.doi.org/#1}{#1}}}

\makeatletter
\@addtoreset{equation}{section}
\makeatother

\begin{document}

\begin{flushright}
\small
IFT-UAM/CSIC-20-181\\
\texttt{arXiv:2012.13323 [hep-th]}\\
February 16\textsuperscript{th}, 2021\\
\normalsize
\end{flushright}

\vspace{1cm}

\begin{center}

  {\Large {\bf The first law of heterotic stringy black hole mechanics\\[.5cm]
      at zeroth order in $\alpha'$}}

\vspace{1cm}

\renewcommand{\thefootnote}{\alph{footnote}}

{\sl\large Zachary Elgood,}\footnote{Email: {\tt zachary.elgood[at]uam.es}}
{\sl\large Dimitrios Mitsios,}\footnote{Email: {\tt  di.mitsios[at]gmail.com}}
{\sl\large Tom\'{a}s Ort\'{\i}n}\footnote{Email: {\tt tomas.ortin[at]csic.es}}
{\sl\large and David Pere\~n\'{\i}guez,}\footnote{Email: {\tt david.perenniguez[at]uam.es}}

\setcounter{footnote}{0}
\renewcommand{\thefootnote}{\arabic{footnote}}

\vspace{1cm}

{\it Instituto de F\'{\i}sica Te\'orica UAM/CSIC\\
C/ Nicol\'as Cabrera, 13--15,  C.U.~Cantoblanco, E-28049 Madrid, Spain}

\vspace{1cm}


{\bf Abstract}
\end{center}
\begin{quotation}
  {\small We re-derive the first law of black hole mechanics in the context of
    the Heterotic Superstring effective action compactified on a torus to
    leading order in $\alpha'$, using Wald's formalism, covariant Lie
    derivatives and momentum maps. The Kalb-Ramond field strength of this
    theory has Abelian Chern-Simons terms which induce Nicolai-Townsend
    transformations of the Kalb-Ramond field.  We show how to deal with all
    these gauge symmetries deriving the first law in terms of manifestly
    gauge-invariant quantities. In presence of Chern-Simons terms, several
    definitions of the conserved charges exist, but the formalism picks up
    only one of them to play a role in the first law. We study explicitly a
    non-extremal, charged, black ring solution of pure $\mathcal{N}=1,d=5$
    supergravity embedded in the Heterotic Superstring effective field theory.

    This work is a first step towards the derivation of the first law at first
    order in $\alpha'$ where, more complicated, non-Abelian, Lorentz
    (``gravitational'') and Yang-Mills Chern-Simons terms are included in the
    Kalb-Ramond field strength. The derivation of a first law is a necessary
    step towards the derivation of a manifestly gauge-invariant entropy
    formula which is still lacking in the literature. In its turn, this
    entropy formula is needed to compare unambiguously macroscopic and
    microscopic black hole entropies.
  }
\end{quotation}

\newpage
\pagestyle{plain}

\tableofcontents


\section{Introduction}

In Ref.~\cite{Wald:1993nt}, Wald showed that, in a theory of gravity invariant
under diffeomorphisms, the black hole entropy is essentially the Noether
charge associated to that invariance. The proof consists in showing that this
charge plays the role of entropy in the first law of black hole mechanics
\cite{Bardeen:1973gs}. In presence of matter, though, some terms in the total
Noether charge are identified with other terms in the first law and only the
``gravitational'' part of the Noether charge can be identified with the
entropy and, in principle, it is necessary to go through the proof of the
first law in order to identify the entropy. In Ref.~\cite{Iyer:1994ys}, Iyer
and Wald studied theories of gravity coupled to matter and found a
prescription (henceforth called the \textit{Iyer-Wald prescription}) to
compute directly the entropy. In the derivation of the Iyer-Wald prescription
though, it was assumed that all the fields of the theory are tensors, a
condition which, in the Standard Model for instance, would only be satisfied
by the metric, since the rest of the fields have some kind of gauge freedom,
including the Higgs ``scalar''. If we decide to describe the gravitational
field through the Vielbein (as the presence of fermions in the Standard Model
demands), not even the gravitational field would be a tensor.

This problem was first noticed by Jacobson and Mohd \cite{Jacobson:2015uqa} in
the context of the theory of General Relativity described by a
Vielbein.\footnote{Fields with gauge freedom had already been correctly dealt
  with in Refs.~\cite{Barnich:2005kq,Compere:2007vx}, for instance.} They
solved the problem by ``improving'' the standard Lie derivative (in the
language of \cite{Frodden:2017qwh,Frodden:2019ylc}) by adding a local Lorentz
transformation that covariantizes it. This Lorentz-covariant Lie derivative,
also known as \textit{Lie-Lorentz derivative} occurs naturally in supergravity
and it was described in that context for arbitrary Lorentz tensors in
Ref.~\cite{Ortin:2002qb}\footnote{See also Ref.~\cite{Ortin:2015hya} and, for
  a more mathematically rigorous point of view, Ref.~\cite{kn:FF}.} building
upon earlier work on the Lie derivative of Lorentz spinors by Lichnerowicz,
Kosmann and others
\cite{kn:Lich,kn:Kos,kn:Kos2,Hurley:cf,Vandyck:ei,Vandyck:gc,Figueroa-O'Farrill:1999va}.
A recent application to supergravity, including the fermion fields can be
found in Ref.~\cite{Aneesh:2020fcr}.

A more general and mathematically rigorous treatment based on the theory of
principal bundles was given in Ref.~\cite{Prabhu:2015vua} by Prabhu, who was
motivated by the problems found by Gao in Ref.~\cite{Gao:2003ys}.  However,
String and Supergravity theories have $p$-form fields with gauge freedom that
cannot be described in that framework. Furthermore, the effective
action and the field strengths often contain Chern-Simons terms which make the
action invariant only up to total derivatives and complicate the gauge
transformations of the $p$-form fields. When the Chern-Simons terms depend on
the spin (Lorentz) connection, gauge invariance and diffeomorphism invariance
become entangled in a very complex form.

One of the simplest theories with a Chern-Simons term in the action is
``minimal'' ($\mathcal{N}=1$) 5-dimensional supergravity
\cite{Cremmer:1980gs}, which only contains a 1-form coupled to gravity. In
order to deal with the lack of exact gauge invariance one has to take into
account the total derivative in the definition of the Noether current
\cite{Barnich:2005kq}.  However, the entropy obtained by this method in
Ref.~\cite{Tachikawa:2006sz} in the case of the ``gravitational'' Chern-Simons
terms (both in the action or in the Kalb-Ramond field strength) of the
Heterotic Superstring effective action turned out to be
gauge-dependent.\footnote{The same happens when one naively uses the Iyer-Wald
  prescription, as noticed in \cite{Elgood:2020xwu,Ortin:2020xdm}.} This
problem was dealt with in Ref.~\cite{Azeyanagi:2014sna}, albeit in a rather
complicated form.

In a recent paper \cite{Elgood:2020svt} we studied the use of gauge-covariant
Lie derivatives in the context of the Einstein-Maxwell theory using momentum
maps to construct the derivatives. Momentum maps arise naturally wherever
symmetries of a base manifold have to be related to gauge transformations
\cite{Ortin:2015hya,Bandos:2016smv} and they are unsurprisingly ubiquitous in
gauged supergravity. As a matter of fact, the Lie-Lorentz derivative can be
constructed in terms of a Lorentz momentum map and in \cite{Elgood:2020svt} we
also used a \textit{Maxwell momentum map} to construct a \textit{Lie-Maxwell
  derivative}, covariant under the gauge transformations of the Maxwell field.

This procedure guarantees the gauge-invariance of the results and, as a
byproduct, we found a very interesting relation between momentum maps and
generalized zeroth laws also observed, in a completely different language by
Prabhu in Ref.~\cite{Prabhu:2015vua}.

In this paper we extend this method to a theory with Abelian Chern-Simons
terms in a field strength: the effective action of the Heterotic Superstring
compactified on a torus to zeroth order in $\alpha'$. This theory can be seen
as a generalization of the theory considered by Comp\`ere in
Ref.~\cite{Compere:2007vx} and as a first step towards dealing with the
effective action of the Heterotic Superstring to first order in $\alpha'$,
which contains non-Abelian and Lorentz (``gravitational'') Chern-Simons terms
of the kind considered by Tachikawa
\cite{Bergshoeff:1988nn,Bergshoeff:1989de}. The introduction of momentum maps
will allow us to obtain invariant results in a rather simple form, basically
because they allow us to determine explicitly the gauge parameters that leave
invariant all the fields of a given solution \cite{Barnich:2001jy}.  They also
allow us to construct forms which are closed on the bifurcation sphere, from
which the definitions of the potentials that appear in the first law will
follow \cite{Copsey:2005se,Compere:2007vx}. The closedness of those forms,
therefore, plays the role of the generalized zeroth law, albeit restricted to
the bifurcation sphere. Hence, we will refer to these properties as the
\textit{restricted generalized zeroth laws}.

As we are going to see in the proof of the first law, there is a very precise,
almost clockwork, relation between the closed forms that satisfy the
restricted generalized zeroth laws and the definitions of the conserved
charges \cite{Regge:1974zd,Abbott:1981ff,Barnich:2001jy,Barnich:2003xg}. Only
when both have been correctly identified is it possible to find the first law
and identify the entropy.

In theories with Chern-Simons terms, several different definitions of charges
have been proposed and used in the literature (see, for instance,
Ref.~\cite{Marolf:2000cb} and references therein). The proof of the first law
demands that we use the so-called \textit{Page charge}, which in this context
is conserved, localized and on-shell gauge invariant. Only when we use this
charge definition for the 1-forms, the closed 1-form associated to the KR
potentials $\Phi^{i}$ over the bifurcation sphere appears
\cite{Copsey:2005se,Compere:2007vx} and the term
$\Phi^{i}\delta \mathcal{Q}_{i}$ of the first law associated to the ``dipole
charges''
\cite{Emparan:2004wy,Copsey:2005se,Astefanesei:2005ad,Rogatko:2005aj,Rogatko:2006xh,Compere:2007vx}
can be identified.

In theories with ``gravitational'' Chern-Simons terms, such as the effective
action of the Heterotic Superstring at first order in $\alpha'$ the same
mechanism should play a role in the proof of the first law, but the terms
that modify the gravitational charges will contribute to the entropy instead
\cite{Elgood:2020nls}. It is in this precise sense that this work is a first step
towards the proof of the first law and the determination of a gauge-invariant
entropy formula for that theory. The previous discussion should have made
clear that such a formula is not yet available, as we have also explained in
Refs.~\cite{Elgood:2020xwu,Ortin:2020xdm}. Even though the calculations of
some black-hole entropies using the Iyer-Wald prescription seem to give the
right value of the entropy in some cases,\footnote{In Ref.~\cite{Cano:2019ycn}
  it was shown that the entropy of the $\alpha'$-corrected non-extremal
  Reissner-Nordstr\"om black hole based in the string embedding of
  Ref.~\cite{Khuri:1995xq}, computed with the entropy formula derived in
  Ref.~\cite{Elgood:2020xwu} using the Iyer-Wald prescription satisfies the
  thermodynamic relation $\partial S/\partial M = T^{-1}$. That entropy
  formula is not invariant under Lorentz transformations, though. In a general
  frame it will give wrong values for the entropy and the reason why it gives
  the right value in that particular case, in the particular frame in which
  the calculation was carried out, sill needs to be explained
  \cite{Elgood:2020nls}. The same entropy formula has been used to compute the entropy
  of some $\alpha'$-corrected extremal black holes and the results, although
  reasonable, cannot be tested using the same relation.}  it is clear that the
results obtained using an entropy formula which is not gauge-invariant cannot
be trusted in general. It is also clear that the comparison between entropies
computed through macroscopic and microscopic methods \cite{Strominger:1996sh}
only make sense if both computations are reliable, and furthermore, only if
the relation between the parameters of the black hole solution and of the
microscopic theory is well understood. At first order in $\alpha'$, there is
no full-proof entropy formula, as we have explained, and the identification of
the parameters of the black-hole solutions (charges) with the numbers of
branes and other parameters that appear in the microscopic entropy, has issues
that still have not been fully understood \cite{Faedo:2019xii}. This is one of
the main motivations for this work.

This paper is organized as follows: in Section~\ref{sec-actionetc} we
introduce the effective action of the Heterotic Superstring compactified on a
torus at leading order in $\alpha'$. In Section~\ref{sec-variations} we study
the action of the symmetries of the theory on the fields, the parameters of
the transformations that leave all of them invariant, and compute the
associated conserved charges, including the Wald-Noether charge. In
Section~\ref{sec-zeroth} we study the restricted generalized zeroth laws that
we will use in the proof of the first law in Section~\ref{sec-firstlaw}. In
Section~\ref{sec-example} we consider as an example the charged, non-extremal,
5-dimensional black ring solution of pure $\mathcal{N}=1,d=5$ supergravity of
Ref.~\cite{Elvang:2004xi} and compute its momentum
maps. Section~\ref{sec-discussion} contains a brief discussion of our results.
In the appendix we show how the Heterotic Superstring effective action
compactified on T$^{4}\times$S$^{1}$ (trivial compactification on T$^{4}$) can
be understood as a model $\mathcal{N}=1,d=5$ supergravity coupled to two
vector supermultiplets, which provides an embedding of this model into the
Heterotic Superstring effective action. We also show how this model can be
consistently trunctated to pure $\mathcal{N}=1,d=5$ supergravity. Again, this
provides an embedding of pure $\mathcal{N}=1,d=5$ supergravity and, in
particular of the black ring solution of Ref.~\cite{Elvang:2004xi} into the
Heterotic Superstring effective action, so we can apply the formulae and
results obtained in the main body of the paper to that solution.

\section{The Heterotic Superstring effective action on  T$^{n}$
  at zeroth order in $\alpha'$}
\label{sec-actionetc}

When the effective action of the Heterotic Superstring at leading order in
$\alpha'$ is compactified on a T$^{n}$, it describes the dynamics of the
$(10-n)$-dimensional (string-frame) metric $g_{\mu\nu}$, Kalb-Ramond 2-form
$B_{\mu\nu}$, dilaton field $\phi$, Kaluza-Klein (KK) and winding 1-forms
$A^{m}{}_{\mu}$ and $B_{m\, \mu}$, respectively, and the scalars that
parametrize the O$(n,n)/$O$(n)\times$O$(n)$ coset space, collected in the
symmetric O$(n,n)$ matrix $M$ that we will write with upper O$(n,n)$ indices
$I,J,\ldots$ as $M^{IJ}$. This means that $M$ satisfies

\begin{equation}
  \label{eq:Onnproperty}
 M^{IJ}\Omega_{JK} M^{KL} \Omega_{LM} = \delta^{I}{}_{M}\, ,  
\end{equation}

\noindent
where

\begin{equation}
\left(\Omega_{IJ} \right)
  \equiv
  \left(
    \begin{array}{cc}
      0 & \mathbb{1}_{n\times n} \\
      \mathbb{1}_{n\times n} & 0 \\
    \end{array}
    \right)\, ,
\end{equation}

\noindent
is the off-diagonal form of the O$(n,n)$ metric. Eq.~(\ref{eq:Onnproperty})
implies that

\begin{equation}
M_{IJ} \equiv (M^{-1})_{IJ} = \Omega_{IK}M^{KL}\Omega_{LJ}\, .  
\end{equation}

Using the notation and conventions of Refs.~\cite{Ortin:2015hya,Ortin:2020xdm}
(in particular, for differential forms, we use those of
Ref.~\cite{Elgood:2020svt}), and calling the physical scalars in $M_{IJ}$
$\phi^{x}$, the action of the $d=(10-n)$-dimensional takes the form

\begin{equation}
\label{eq:heterotic(10-n)order0}
\begin{aligned}
  S[e^{a},B,\phi,\mathcal{A}^{I},\phi^{x}]
  & =
  \frac{g_{s}^{(d)\, 2}}{16\pi G_{N}^{(d)}} \int e^{-2\phi}
  \left[ (-1)^{d-1} \star (e^{a}\wedge e^{b}) \wedge R_{ab}
    -4d\phi\wedge \star d\phi
  \right.
  \\
  & \\
  & \hspace{.5cm}
  \left.
      -\tfrac{1}{8}dM_{IJ}\wedge \star dM^{IJ}
    +(-1)^{d}\tfrac{1}{2}M_{IJ}\mathcal{F}^{I}\wedge \star \mathcal{F}^{J}
    +\tfrac{1}{2}H\wedge \star H
  \right]
  \\
  & \\
  & \equiv
  \int  \mathbf{L}\, .
\end{aligned}
\end{equation}

\noindent
In this action $e^{a}=e^{a}{}_{\mu}dx^{\mu}$ are the string-frame
Vielbeins, $\star$ stands for the Hodge dual and, therefore

\begin{equation}
  \star (e^{a}\wedge e^{b})
  =
  \frac{1}{(d-2)!}\epsilon_{c_{1}\cdots c_{d-2}}{}^{ab}
  e^{c_{1}}\wedge \cdots \wedge e^{c_{d-2}}\, .  
\end{equation}

\noindent
Furthermore, $\omega^{ab}=\omega_{\mu}{}^{ab}dx^{\mu}$ is the Levi-Civita spin
connection\footnote{It is antisymmetric $\omega^{ab}=-\omega^{ba}$ and
  satisfies $De^{a}= de^{a}-\omega^{a}{}_{b}\wedge e^{b}=0$. We are using the
  second-order formalism.} and
$R^{ab}= \tfrac{1}{2}R_{\mu\nu}{}^{ab}dx^{\mu}\wedge dx^{\nu}$ is its field
strength (the curvature) 2-form, defined as

\begin{equation}
  \label{eq:curvaturedefined}
  R^{ab} \equiv d\omega^{ab} -\omega^{a}{}_{c}\wedge \omega^{cb}\, .
\end{equation}

\noindent
$g^{(d)}_{s}$ and $G_{N}^{(d)}$ are, respectively, the
$d=(10-n)$-dimensional string coupling and Newton
constant. \footnote{They are related to the 10-dimensional constants
  through the volume of the T$^{n}$, $V_{n}$, by
  \begin{subequations}
     \begin{align}
       g_{s}^{2}
       & =
         V_{n}/(2\pi\ell_{s})^{n}g_{s}^{(d)\, 2}\, ,
       \\
                 & \nonumber \\
     \label{eq:relationsbetweenconstants}
       G_{N}^{(10)}
       & =
         G_{N}^{(d)} V_{n}\, . 
     \end{align}
   \end{subequations}
}

\noindent
$\mathcal{F}^{I}$ is the O$(n,n)$ vector of the 2-form field strengths of the
KK and winding vectors

\begin{equation}
    \label{eq:Fdef}
    \mathcal{F}^{I}
  \equiv
  \left(
    \begin{array}{c}
      F^{m} \\ G_{m} 
    \end{array}
  \right)\, ,
  \hspace{.5cm}
  F^{m} = dA^{m}\, ,
  \hspace{.5cm}
  G_{m}  = dB_{m}\, ,
\end{equation}

\noindent
which can also be defined in terms of the O$(n,n)$ vector of 1-forms denoted
by $\mathcal{A}^{I}$

\begin{equation}
  \mathcal{A}^{I}
  \equiv
  \left(
    \begin{array}{c}
      A^{m} \\ B_{m} \\
    \end{array}
  \right)\, ,
  \hspace{1cm}
  \mathcal{F}^{I}
  =
  d\mathcal{A}^{I}\, .
\end{equation}

\noindent
$H$ is the Kalb-Ramond 3-form field strength, defined by 

\begin{equation}
    \label{eq:Hdef}
    H
    \equiv
    dB -\tfrac{1}{2}\mathcal{A}_{I}\wedge d\mathcal{A}^{I}\, ,
    \hspace{1cm}
    \mathcal{A}_{I} = \Omega_{IJ}\mathcal{A}^{J}\, .
\end{equation}

The kinetic term of the scalars $\phi^{x}$ that parametrize the
O$(n,n)/($O$(n)\times$O$(n))$ coset space can also be written in the form

\begin{equation}
  -\tfrac{1}{8}dM_{IJ}\wedge \star dM^{IJ}
  =
  \tfrac{1}{2}g_{xy}d\phi^{x}\wedge \star d\phi^{y}\, ,
\end{equation}

\noindent
where the metric $g_{xy}(\phi)$ is given by

\begin{equation}
g_{xy} \equiv \tfrac{1}{4}\left(\partial_{x}M_{IK}M^{KJ}\right)
\left(\partial_{y}M_{JK}M^{KI}\right)\, .
\end{equation}

Under a general variation of the fields, the action varies as

\begin{equation}
  \label{eq:deltaS}
  \begin{aligned}
    \delta S
    & =
    \int \left\{ \mathbf{E}_{a}\wedge \delta e^{a}
      +\mathbf{E}_{B}\wedge \delta B
      +\mathbf{E}_{\phi}\delta\phi +\mathbf{E}_{I}\wedge \delta \mathcal{A}^{I}
      +\mathbf{E}_{x}\delta \phi^{x}
      +d\mathbf{\Theta}(\varphi,\delta\varphi) \right\}\, ,
\end{aligned}
\end{equation}

\noindent
where, suppressing the factors of $g^{(d)\, 2}(16\pi G_{N}^{(d)})^{1}$ for
simplicity, the Einstein equations $\mathbf{E}_{a}$ are given by

\begin{equation}
  \begin{aligned}
    \label{eq:Ea}
  \mathbf{E}_{a}
  & =
    e^{-2\phi} \imath_{a}\star (e^{c}\wedge e^{d})\wedge R_{cd}
    -2\mathcal{D}(\imath_{b}de^{-2\phi})\wedge \star(e^{b}\wedge e^{c})g_{ca}
    \\
    & \\
    &
      \hspace{.5cm}
      +(-1)^{d-1}4e^{-2\phi}
      \left(\imath_{a}d\phi \star d\phi+d\phi\wedge \imath_{a}\star d\phi\right)
    \\
    & \\
    &
      \hspace{.5cm}
      +\frac{(-1)^{d}}{2}e^{-2\phi}g_{xy}
      \left(\imath_{a}d\phi^{x} \star d\phi^{y}+d\phi^{x}\wedge \imath_{a}\star d\phi^{y}\right)
    \\
    & \\
    &
       \hspace{.5cm}
      +\frac{1}{2}e^{-2\phi}M_{IJ}\left(\imath_{a}\mathcal{F}^{I}\wedge \star \mathcal{F}^{J}
      -\mathcal{F}^{I}\wedge \imath_{a}\star \mathcal{F}^{J}
      \right)
     \\
    & \\
    &
      \hspace{.5cm}
      +\frac{(-1)^{d}}{2}e^{-2\phi}
      \left(\imath_{a}H\wedge \star H+H\wedge \imath_{a}\star H\right)\, ,    
  \end{aligned}
\end{equation}

\noindent
the equations of motion of the matter fields are given by 

\begin{subequations}
  \begin{align}
    \label{eq:EB}
    \mathbf{E}_{B}
    & =
      -d\left(e^{-2\phi}\star H\right)\, ,
          \\
    & \nonumber \\
    \label{eq:Ephi}
    \mathbf{E}_{\phi}
    & =
      8d\left(e^{-2\phi}\star d\phi\right)-2\mathbf{L}\, ,
          \\
    & \nonumber \\
    \label{eq:EI}
    \mathbf{E}_{I}
    & =
    \tilde{\mathbf{E}}_{I}
      +\tfrac{1}{2}\mathbf{E}_{B}\wedge \mathcal{A}_{I}\, ,
          \\
    & \nonumber \\
    \label{eq:tildeEI}
    \tilde{\mathbf{E}}_{I}
    & \equiv
      -\left\{d\left(e^{-2\phi} M_{IJ}\star \mathcal{F}^{J} \right)
      +(-1)^{d-1}e^{-2\phi}\star H\wedge \mathcal{F}_{I}\right\}\, ,
          \\
    & \nonumber \\
    \label{eq:Ex}
    \mathbf{E}_{x}
    & =
      -g_{xy}\left[d\left(e^{-2\phi}\star d\phi^{y} \right)
      +e^{-2\phi}\Gamma_{zw}{}^{y}d\phi^{z}\wedge \star d\phi^{w} \right]
      +\frac{(-1)^{d}}{2} e^{-2\phi} \partial_{x}M_{IJ}\mathcal{F}^{I}\wedge \star
      \mathcal{F}^{J}\, ,
 \end{align}
 \end{subequations}
 
 \noindent
 and

 \begin{equation}
   \label{eq:Theta}
  \begin{aligned}
    \mathbf{\Theta}(\varphi,\delta\varphi)
    & =
    -e^{-2\phi}\star (e^{a}\wedge e^{b})\wedge \delta \omega_{ab}
    +2\imath_{a}de^{-2\phi}\star(e^{a}\wedge e^{b})\wedge \delta e_{b}
    \\
    & \\
    &
    \hspace{.5cm}
    -8e^{-2\phi}\star d\phi\delta\phi
        -\tfrac{1}{4}e^{-2\phi}\star dM^{IJ}\delta M_{IJ}
    \\
    & \\
    &
    \hspace{.5cm}
    +e^{-2\phi}M_{IJ}\star\mathcal{F}^{J}\wedge \delta\mathcal{A}^{I}
    +e^{-2\phi}\star H\wedge \left(\delta B
      +\tfrac{1}{2}\mathcal{A}_{I}\wedge \delta\mathcal{A}^{I}\right)\, .
  \end{aligned}
\end{equation}

The equations of motion of the 1-forms $\mathbf{E}_{I}$ can be written in the
alternative form

\begin{equation}
  \label{eq:EIalt}
  \mathbf{E}_{I}
  =
  -d\left\{
    e^{-2\phi} M_{IJ}\star \mathcal{F}^{J} +\star H\wedge\mathcal{A}_{I}
  \right\}
    -\tfrac{1}{2}\mathbf{E}_{B}\wedge \mathcal{A}_{I}\, .
\end{equation}

\noindent
This form appears naturally in the definition of the electric charges
Eq.~(\ref{eq:QIdef}).

Here, and in what follows, $\varphi$ stands for all the fields of the theory.
$\mathbf{E}_{\varphi}$ denotes collectively all their equations of motion.

\section{Variations of the fields}
\label{sec-variations}

In this section we are going to study the transformations of the fields under
the different symmetries of the action and determine which parameters of the
transformations leave a complete field configuration invariant. The conserved
charges of those configurations will be associated to those parameters. As a
general rule, only if one combines several transformations can one find
parameters that simultaneously leave all the fields invariant.

The simplest case in which this happens will involve the gauge transformations
of the 1-form fields: the parameters that leave them invariant do not leave the KR field
invariant at the same time, unless we perform a KR gauge
transformation with a parameter related to that of the other gauge
symmetry. As a result, there is an additional term in the formula that gives
the electric charges, but it is the presence of this additional term that
guarantees the conservation of the charge and the independence of the integration
surface (as long as we do not include sources, that is, on-shell).

The transformation of several fields under diffeomorphisms must also be
supplemented by ``compensating'' gauge transformations, including local
Lorentz transformations if we want all the fields to be left invariant by
those generating isometries (Killing vectors). There are several ways of
understanding this need but we believe that the most fundamental is to realize
that fields with gauge freedoms (\textit{i.e.}~all fields except for the
metric and the dilaton field) are not tensors and do not transform as such
under diffeomorphisms. The ``compensating gauge transformations'' can be seen
as gauge transformations induced by the diffeomorphisms. Only when they are
properly taken into account can one find Killing vector fields that leave all the fields
invariant. Furthermore, only then the vanishing of the
variations of the fields is invariant under gauge transformations. A more
detailed discussion and additional references to this topic can be found in
Ref.~\cite{Elgood:2020svt}. The conserved charge associated to
diffeomorphisms, the Wald-Noether charge, will therefore include terms
related to gauge symmetries and their associated conserved charges, which
will ultimately contribute to the first law.

As we will see, only when all these details are properly taken
into account can the first law be proven and the entropy identified.

We start by describing the gauge symmetries of the theory (other than
diffeomorphisms) and the associated conserved charges.

\subsection{Gauge transformations}

The gauge transformations of the fields are 

\begin{subequations}
  \label{eq:gaugetransformations}
  \begin{align}
    \label{eq:deltasigmaea}
    \delta_{\sigma}e^{a}
    & =
    \sigma^{a}{}_{b}e^{b}\, ,
    \\
    & \nonumber \\
    \delta_{\chi}\mathcal{A}^{I}
    & =
    d\chi^{I}\, ,
    \\
    & \nonumber \\
    \delta B
    & =
      (\delta_{\Lambda}+\delta_{\chi}) B
      =
      d\Lambda
      +\tfrac{1}{2}\chi_{I}d\mathcal{A}^{I}\, , 
  \end{align}
\end{subequations}

\noindent
where $\sigma^{(ab)}(x)=0$ are the parameters of local Lorentz
transformations, $\chi^{I}(x)$ is a O$(n,n)$ vector if scalar gauge parameters
and $\Lambda=\Lambda_{\mu}(x)dx^{\mu}$ is a 1-form gauge parameter. They leave
invariant the field strengths $\mathcal{F}^{I}$ and $H$, but they induce the
following transformations on the spin connection and curvature

\begin{subequations}
  \begin{align}
    \label{eq:deltasigmaomega}
\delta_{\sigma} \omega^{ab}
    & =
      \mathcal{D}\sigma^{ab}
      =
       d\sigma^{ab}-2\omega^{[a|}{}_{c}\sigma^{c|b]}\, ,
    \\
    & \nonumber \\
      \delta_{\sigma} R^{ab}
      & =
        2\sigma^{[a|}{}_{c}R^{c|b]}\, .
  \end{align}
\end{subequations}

For the sake of completeness and later use, we quote the Ricci identity in our
conventions:

\begin{equation}
  \label{eq:Ricci}
  \mathcal{D}\mathcal{D}\sigma^{ab}
   =
    -2R^{[a|}{}_{c}\sigma^{c|b]}
    =
    \delta_{\sigma}R^{ab}\, .
\end{equation}

The action is manifestly invariant under these gauge transformations. This
leads to the following Noether identities 

\begin{subequations}
  \begin{align}
    \label{eq:noether1}
\mathbf{E}^{[a}\wedge e^{b]}
  & =
    0\, ,
  \\
  & \nonumber \\
    \label{eq:noether2}
  d\tilde{\mathbf{E}}_{I} +(-1)^{d}\mathbf{E}_{B}\wedge \mathcal{F}_{I}
  & =
0\, ,
  \\
    & \nonumber \\
        \label{eq:noether3}
      d\mathbf{E}_{B}
    & = 0\, ,
\end{align}
\end{subequations}

\subsection{Gauge charges}

Let us study the conserved charges associated to the gauge transformations
$\delta_{\chi},\delta_{\Lambda}$ and, for the sake of completeness,
$\delta_{\sigma}$, starting with $\delta_{\Lambda}$, which is simpler to deal
with.

The variation of the action under $\delta_{\Lambda}$ transformations follows
from Eqs.~(\ref{eq:deltaS}) and (\ref{eq:Theta})

\begin{equation}
  \label{eq:deltaSLambda}
  \begin{aligned}
    \delta_{\Lambda} S
    & =
    \int \left\{
      \mathbf{E}_{B}\wedge \delta_{\Lambda} B
      +d\left(e^{-2\phi}\star H\wedge \delta_{\Lambda} B \right) \right\}
    \\
    & \\
    & =
        \int \left\{
      \mathbf{E}_{B}\wedge d\Lambda
      +d\left(e^{-2\phi}\star H\wedge d\Lambda \right) \right\}\, .
\end{aligned}
\end{equation}

\noindent
Integrating by parts the first term and using the Noether identity
Eq.~(\ref{eq:noether3})

\begin{equation}
  \label{eq:deltaSLambda2}
    \delta_{\Lambda} S
     =
    \int d\left(\Lambda\wedge \mathbf{E}_{B}
      +e^{-2\phi}\star H\wedge d\Lambda \right)
    \equiv
    \int d\mathbf{J}[\Lambda]\, .
\end{equation}

\noindent
The invariance of the action under these gauge transformations indicates that
the current $\mathbf{J}[\Lambda]$ must be locally exact, so that, locally,
there is a $\mathbf{Q}[\Lambda]$ such that
$\mathbf{J}[\Lambda] = d\mathbf{Q}[\Lambda]$. It is easy to see that

\begin{equation}
\mathbf{Q}[\Lambda] = \Lambda\wedge \left(e^{-2\phi}\star H \right)\, .
\end{equation}

The conserved charge is given by the integral of the conserved
$(d-2)$-form $\mathcal{Q}[\Lambda]$ over $(d-2)$-dimensional compact
surfaces $\mathcal{S}_{d-2}$ for $\Lambda$s that leave invariant the
KR field $B$s. These are closed 1-forms. Following
\cite{Copsey:2005se,Compere:2007vx}, using the Hodge decomposition
theorem, these closed 1-forms $\Lambda$ can be written as the sum of
an exact and a harmonic form $\Lambda_{e}=d\lambda$ and $\Lambda_{h}$,
respectively. The exact form $\Lambda_{e}$ will not contribute to the
integral on-shell because

\begin{equation}
  Q(\Lambda_{e})
  =
  \int_{\mathcal{S}_{d-2}}d\lambda \wedge \left(e^{-2\phi}\star H \right)
  =
  \int_{\mathcal{S}_{d-2}}d\left[\lambda \wedge \left(e^{-2\phi}\star H \right)\right]
  -\int_{\mathcal{S}_{d-2}}\lambda \wedge \mathbf{E}_{B}\, .
\end{equation}

\noindent
Therefore, 

\begin{equation}
  \label{eq:QLambdacharge-0}
  Q(\Lambda) = \int_{\mathcal{S}_{d-2}}
  \Lambda_{h}\wedge \left(e^{-2\phi}\star H \right)\, .
\end{equation}

\noindent
Then, using the duality between homology and cohomology, if $C_{\Lambda_{h}}$
is the $(d-3)$-cycle dual to $\Lambda_{h}$, we arrive at the charges

\begin{equation}
  \label{eq:QLambdacharge}
  Q(\Lambda)
  =
  -\frac{g^{(d)\, 2}_{s}}{16\pi G_{N}^{(d)}}\int_{C_{\Lambda_{h}}}e^{-2\phi}\star H\, ,
\end{equation}

\noindent
where we have added a conventional sign and recovered the factor of
$g^{(d)\, 2}_{s}(16\pi G_{N}^{(d)})^{-1}$ that we have omitted. From the
string theory point of view, these charges are just winding numbers of strings
whose transverse space is the cycle $C_{\Lambda_{h}}$. Two homologically
equivalent cycles give the same value of the charge on-shell, that is, if
there are no sources of the KR field in the $(d-2)$-dimensional volume whose
boundary is the union of the two properly oriented $(d-3)$-cycles.

Let us now consider the conserved charges associated to the invariance under
$\delta_{\chi}$. This transformation acts on the 1-forms $\mathcal{A}^{I}$ and
on the KR 2-form $B$. Transformations with constant $\chi^{I}$ (closed
$0$-forms) leave invariant the 1-forms, but they do not leave invariant $B$.
They only change it by an exact 2-form
$d\left(\tfrac{1}{2}\chi_{I}\mathcal{A}^{I}\right)$. Thus, we must add a
compensating $\Lambda$ gauge transformation with parameter
$\Lambda_{\chi}= -\tfrac{1}{2}\chi_{I}\mathcal{A}^{I}$ and consider the
transformation of $B$

\begin{equation}
  \label{eq:deltachiBmodified}
  \delta_{\chi}B
  =
  -\tfrac{1}{2}d\left(\chi_{I}\mathcal{A}^{I}\right)
  +\tfrac{1}{2}\chi_{I}d\mathcal{A}^{I}
  =
  -\tfrac{1}{2}d \chi_{I}\wedge \mathcal{A}^{I}\, .
\end{equation}

\noindent
Then,  from Eqs.~(\ref{eq:deltaS}) and (\ref{eq:Theta}) and the
modified transformation rule Eq.~(\ref{eq:deltachiBmodified}), we get

\begin{equation}
  \label{eq:deltaSchi}
  \begin{aligned}
    \delta_{\chi} S
    & =
    \int \left\{
      \mathbf{E}_{B}\wedge \delta_{\chi} B
      +\mathbf{E}_{I}\wedge \delta_{\chi} \mathcal{A}^{I}
          \right.
    \\
    & \\
    &
    \hspace{.5cm}
    \left.
      +d\left[e^{-2\phi}M_{IJ}\star\mathcal{F}^{J}\wedge \delta_{\chi}\mathcal{A}^{I}
    +e^{-2\phi}\star H\wedge \left(\delta_{\chi} B
      +\tfrac{1}{2}\mathcal{A}_{I}\wedge \delta_{\chi}\mathcal{A}^{I}\right)\right]
      \right\}\, ,
      \\
    & \\
    & =
    \int \left\{
      \left(\mathbf{E}_{I}
      +\tfrac{1}{2}\mathbf{E}_{B}\wedge \mathcal{A}_{I} \right)\wedge d \chi^{I}
      +d\left[\left(e^{-2\phi}M_{IJ}\star\mathcal{F}^{J}
    +e^{-2\phi}\star H\wedge \mathcal{A}_{I}\right)\wedge d \chi^{I}\right]
      \right\}\, .
\end{aligned}
\end{equation}

\noindent
Integrating by parts the first term and using the Noether identities
Eqs.~(\ref{eq:noether2}) and (\ref{eq:noether3}) we get

\begin{equation}
  \label{eq:deltaSchi2}
  \begin{aligned}
    \delta_{\chi} S
    & =
    \int d\left\{
    (-1)^{d-1} \chi^{I}\left(\mathbf{E}_{I}
      +\tfrac{1}{2}\mathbf{E}_{B}\wedge \mathcal{A}_{I} \right)
      +\left(e^{-2\phi}M_{IJ}\star\mathcal{F}^{J}
    +e^{-2\phi}\star H\wedge \mathcal{A}_{I}\right)\wedge d \chi^{I}
      \right\}\, .
\end{aligned}
\end{equation}

The usual argument leads to the conserved $(d-2)$-form

\begin{equation}
  \mathbf{Q}[\chi]
  =
(-1)^{d}\chi^{I}\left(e^{-2\phi}M_{IJ}\star\mathcal{F}^{J}
    +e^{-2\phi}\star H\wedge \mathcal{A}_{I}\right)\, ,
\end{equation}

\noindent
and the definition of electric charges

\begin{equation}
  \label{eq:QIdef}
  \mathcal{Q}_{I}
  =
  \frac{(-1)^{d-1}g^{(d)\, 2}_{s}}{16\pi G_{N}^{(d)}} \int_{\mathcal{S}_{(d-2)}}
  \left(e^{-2\phi}M_{IJ}\star\mathcal{F}^{J}
    +e^{-2\phi}\star H\wedge \mathcal{A}_{I}\right)\, ,
\end{equation}

\noindent
where we have added a conventional sign. Again, this charge is
on-shell invariant under homologically-equivalent deformations of
$\mathcal{S}_{(d-2)}$. This follows from the equation of motion
written in the alternative form Eq.~(\ref{eq:EIalt}). It is also
on-shell invariant under the $\delta_{\chi}$ transformations, in spite
of the explicit occurrence of the vector fields $\mathcal{A}_{I}$: the
second term in the integrand has the same structure as the integrand
of the KR charge and, for the same reason, it is invariant on-shell
when we add to $\mathcal{A}_{I}$ exact 1-forms.

This charge is, in the terminology used by Marolf in
Ref.~\cite{Marolf:2000cb}, a \textit{Page charge} but, as we have
explained, apart from localized and conserved, it is also gauge
invariant on-shell. The formalism leads us to use precisely this
charge, which will be the one occurring in the first law of black hole
mechanics.

Finally, let us consider the charge associated to the invariance under
local Lorentz transformations $\delta_{\sigma}$, which act on the
Vielbein and on all the fields derived from it: spin connection and
curvature. Let us postpone for the time being the conditions that the
parameters that leave all of them invariant have to satisfy and lets
study the transformation of the action. From
Eqs.~(\ref{eq:deltaS}) and (\ref{eq:Theta}) we find

\begin{equation}
  \label{eq:deltasigmaS}
  \begin{aligned}
    \delta_{\sigma} S
    & =
    \int \left\{ \mathbf{E}_{a}\wedge \delta_{\sigma} e^{a}
      +d\left[
            -e^{-2\phi}\star (e^{a}\wedge e^{b})\wedge \delta_{\sigma}\omega_{ab}
    +2\imath_{a}de^{-2\phi}\star(e^{a}\wedge e^{b})\wedge \delta_{\sigma} e_{b}
        \right]\right\}\, ,
\end{aligned}
\end{equation}

\noindent
and using Eqs.~(\ref{eq:deltasigmaea}) and (\ref{eq:deltasigmaomega})
and the Noether identity Eq.~(\ref{eq:noether1}), we find that the
integrand immediately reduces to a total derivative,

\begin{equation}
  \label{eq:deltasigmaS-2}
  \begin{aligned}
    \delta_{\sigma} S
    & =
    \int d\mathbf{J}[\sigma]\, ,
    \\
    & \\
    \mathbf{J}[\sigma]
    & =
            (-1)^{d-1}e^{-2\phi}\mathcal{D}\sigma_{ab}\wedge\star (e^{a}\wedge e^{b})
    +2\sigma_{bc}\imath_{a}de^{-2\phi}\star(e^{a}\wedge e^{b})\wedge e^{c}\, .
\end{aligned}
\end{equation}

\noindent
The standard argument tells us that
$\mathbf{J}[\sigma]=d \mathbf{Q}[\sigma]$. Integrating by parts the first term

\begin{equation}
    \mathbf{J}[\sigma]
    =
            d\left\{(-1)^{d-1}e^{-2\phi}\sigma_{ab}\star (e^{a}\wedge e^{b})\right\}
            +3\left(\sigma_{[bc}\imath_{a]}de^{-2\phi}\right)
            \star(e^{a}\wedge e^{b})\wedge e^{c}\, .
\end{equation}

\noindent
The last term vanishes identically because\footnote{Here we use the property
  \begin{equation}
  \star \omega^{(p)} \wedge \hat{\xi} = \star \imath_{\xi}\omega^{(p)}\, ,  
  \end{equation}
  which is valid for any $p$-form $\omega^{(p)}$ and any vector field
  $\xi=\xi^{\mu}\partial_{\mu}$ and its dual 1-form
  $\hat{\xi}=\xi_{\mu}dx^{\mu}$.}
$\star(e^{a}\wedge e^{b})\wedge e^{c}= 2\eta^{c[a}\star e^{b]}$ and we arrive
at

\begin{equation}
  \label{eq:Lorentzconservedd-2form}
  \mathbf{Q}[\sigma]
  =
  (-1)^{d-1}e^{-2\phi}\star (e^{a}\wedge e^{b})\wedge\sigma_{ab}\, .
\end{equation}

Now we have to consider Lorentz parameters that leave all the
fields invariant. The spin connection and curvature are left invariant by
covariantly constant parameters

\begin{equation}
\mathcal{D}\sigma^{a}{}_{b}=0\, ,  
\end{equation}

\noindent
but the invariance of the Vielbein $\sigma^{a}{}_{b}e^{b}=0$ can only
be satisfied for $\sigma^{a}{}_{b}=0$, and would
automatically imply the vanishing of $\mathbf{Q}[\sigma]$.

The $(d-2)$-form, though, reappears in the proof of the first law for
a Lorentz parameter that is covariantly constant over the bifurcation
surface. We also notice that terms of higher order in the Lorentz
curvature, such as those which arise with $\alpha'$ corrections, lead
to a non-vanishing Lorentz charge Ref.~\cite{Elgood:2020nls}
.

\subsection{Diffeomorphisms and covariant Lie derivatives}

As we have discussed in the introduction, out of the fundamental fields of our
theory, only the dilaton $\phi$ and the O$(n,n)/($O$(n)\times$O$(n))$ scalars
$\phi^{x}$ transform as a tensor under diffeomorphisms
$\delta_{\xi} x^{\mu} = \xi^{\mu}$, that is\footnote{The metric
  $g_{\mu\nu} = \eta_{ab}e^{a}{}_{\mu} e^{b}{}_{\mu}$ and the 2- and 3-form
  field strengths $\mathcal{F},H$ also transform as tensors:
  \begin{subequations}
    \begin{align}
          \delta_{\xi} g_{\mu\nu}
    & =
      -\pounds_{\xi}g_{\mu\nu} = -2\nabla_{(\mu}\xi_{\nu)}\, ,
      \\
      & \nonumber \\
      \delta_{\xi} \mathcal{F}
      & =
        -\pounds_{\xi}\mathcal{F}
        =
        -(\imath_{\xi}d +d\imath_{\xi})\mathcal{F}\, ,
      \\
      & \nonumber \\
      \delta_{\xi} H
      & =
        -\pounds_{\xi}H
        =
        -(\imath_{\xi}d +d\imath_{\xi})H\, .
    \end{align}
  \end{subequations}
}

\begin{subequations}
  \begin{align}
    \delta_{\xi} \phi
     & =
       -\pounds_{\xi}\phi = -\imath_{\xi}d\phi\, ,
    \\
    & \nonumber \\
    \delta_{\xi} \phi^{x}
     & =
       -\pounds_{\xi}\phi^{x} = -\imath_{\xi}d\phi^{x}\, .
  \end{align}
\end{subequations}

The Vielbein $e^{a}$, the vectors (1-forms), $\mathcal{A}$, and the KR 2-form,
$B$, have gauge freedoms and transform as tensors up to \textit{compensating}
gauge transformations. These compensating gauge transformations can be
determined by

\begin{enumerate}
\item Requiring gauge-covariance of the complete transformation law (which
  can then be interpreted as a gauge-covariant Lie derivative) and

\item Imposing that, for diffeomorphisms which are symmetries of the field
  configuration that we are considering (in particular, for isometries), the
  complete transformation (covariant Lie derivative) vanishes. The first
  condition ensures that this vanishing is gauge-invariant.
\end{enumerate}

In what follows we will denote by $k$ the vector fields $\xi$ that
generate diffeomorphisms that leave invariant the complete field
configuration. $k$ is, in particular, a Killing vector of the metric.

In a recent paper \cite{Elgood:2020svt} we reviewed the construction of a Lie
derivative of the Vielbein, spin connection and curvature covariant under
local Lorentz transformations (\textit{Lie-Lorentz derivative}) of
Refs.~\cite{Ortin:2002qb,Ortin:2015hya} that build upon earlier work by
Lichnerowicz, Kosmann and others \cite{kn:Lich,kn:Kos,kn:Kos2,Hurley:cf}. In
Ref.~\cite{Elgood:2020svt} we also dealt with Abelian vector fields in similar
terms. It is convenient to quickly review these results starting with the
Abelian vector case, adapted to the present situation.

The transformation of the  Abelian vector fields $\mathcal{A}^{I}$ under
diffeomorphisms can be defined as

\begin{equation}
\delta_{\xi}\mathcal{A}^{I} = - \mathbb{L}_{\xi}\mathcal{A}^{I}\, ,
\end{equation}

\noindent
where $\mathbb{L}_{\xi}\mathcal{A}^{I}$ is the \textit{Lie-Maxwell derivative},
defined by 

\begin{equation}
  \label{eq:LieMaxwelldef}
\mathbb{L}_{\xi}\mathcal{A}^{I} \equiv \imath_{\xi}\mathcal{F}^{I} +d\mathcal{P}_{\xi}{}^{I}\, .
\end{equation}

\noindent
Here $\mathcal{P}_{\xi}{}^{I}$ is a gauge-invariant O$(n,n)$ vector of
functions that depends on $\mathcal{A}^{I}$ and on the generator of
diffeomorphisms $\xi$ and it is assumed to have the property that,
when $\xi=k$, it satisfies the equation

\begin{equation}
\label{eq:Pkdef}
d\mathcal{P}_{k}{}^{I}=- \imath_{k}\mathcal{F}^{I}\, .
\end{equation}

The invariance of the 2-form $\mathcal{F}^{I}$ guarantees the local
existence of $\mathcal{P}_{k}{}^{I}$, which is known as the
\textit{momentum map} associated to $k$. On the other hand,
Eq.~(\ref{eq:Pkdef}) ensures that the two properties of the variations
of the fields under diffeomorphisms that we have demanded are
satisfied. Finally, observe that the Lie-Maxwell derivative is just a
combination of the standard Lie derivative plus a compensating gauge
transformation with parameter

\begin{equation}
  \label{eq:LMparameter}
\chi_{\xi}{}^{I}= \imath_{\xi}\mathcal{A}^{I} -\mathcal{P}_{\xi}{}^{I}\, .
\end{equation}

For fields with Lorentz indices (Vielbein, spin connection and
curvature), the variation under diffeomorphisms is also given by
(minus) a Lorentz-covariant generalization of the Lie derivative
$\delta_{\xi} = - \mathbb{L}_{\xi}$ usually called \textit{Lie-Lorentz
  derivative}
Refs.~\cite{Ortin:2002qb,Ortin:2015hya,kn:Lich,kn:Kos,kn:Kos2,Hurley:cf}.
This derivative can also be constructed by adding to the standard Lie
derivative a compensating Lorentz transformation with the parameter

\begin{equation}
  \label{eq:LLparameter}
  \sigma_{\xi}{}^{ab}
  =
  \imath_{\xi}\omega^{ab} -\nabla^{[a}\xi^{b]}\, .
\end{equation}

For the Vielbein, the Lie-Lorentz derivative can be expressed in several
equivalent and manifestly Lorentz-covariant forms

\begin{subequations}
  \begin{align}
  \label{eq:LLeam}
  \mathbb{L}_{\xi}e^{a}{}_{\mu}
   & =
  \tfrac{1}{2}e^{a\,
     \nu}\left(\nabla_{\mu}\xi_{\nu}+\nabla_{\nu}\xi_{\mu}\right)
    \\
    & \nonumber \\
      \label{eq:LLeam2}
  \mathbb{L}_{\xi}e^{a}
    & =
      \mathcal{D}\xi^{a}+P_{\xi}{}^{a}{}_{b}e^{b}\, ,
  \end{align}
\end{subequations}

\noindent
where

\begin{equation}
  \label{eq:Pxiab}
  P_{\xi}{}^{ab} \equiv \nabla^{[a}\xi^{b]}\, ,
\end{equation}

\noindent
satisfies, when $\xi=k$, the equation

\begin{equation}
\imath_{k}R^{ab} = -\mathcal{D}P_{k}{}^{ab}\, ,
\end{equation}

\noindent
that shows that we can view $P_{k}{}^{ab}$ as a momentum map as
well.\footnote{Compare this equation to Eq.~(\ref{eq:Pkdef}).}

In the form Eq.~(\ref{eq:LLeam}) we immediately see that the
Lie-Lorentz derivative of the Vielbein vanishes when $\xi=k$, a
Killing vector. The same is true for the connection and curvature.

Observe that $P_{\xi}{}^{ab}$ transforms covariantly under local Lorentz
transformations.

The above transformation of the Vielbein induce the following transformations
of the spin connection and curvature that we quote for later use:

\begin{subequations}
  \begin{align}
    \label{eq:deltaxiomegaab}
    \delta_{\xi}\omega^{ab}
    & =
      -\mathbb{L}_{\xi}\omega^{ab}
  =
    -\left(  \imath_{\xi}R^{ab} +\mathcal{D}P_{\xi}{}^{ab}\right)\, ,
    \\
    & \nonumber \\
    \delta_{\xi}R^{ab}
    & =
      -\mathbb{L}_{\xi}R^{ab}
  =
      -\left(\mathcal{D}\imath_{\xi}R^{ab}
      -2P_{\xi}{}^{[a}{}_{c}R^{b]c}\right)\, .
  \end{align}
\end{subequations}

Observe that the Lie-Lorentz derivative of the spin connection has the
same structure as that of the Abelian connection $\mathcal{A}^{I}$ in
Eq.~(\ref{eq:LieMaxwelldef}), \textit{i.e.}~the inner product of $\xi$
with the curvature plus the derivative of the momentum map.

In asymptotically-flat stationary black-hole spacetimes with
bifurcate horizon, if $k$ is the Killing vector whose Killing horizon
coincides with the event horizon and $\mathcal{BH}$ is the bifurcation
sphere, 

\begin{equation}
  \label{eq:Pkab}
  P_{k}{}^{ab}
  =
  \nabla^{[a}k^{b]} \stackrel{\mathcal{BH}}{=} \kappa n^{ab}\, ,  
\end{equation}

\noindent
where $\kappa$ is the surface gravity and $n^{ab}$ is the binormal to
the event horizon, with the normalization $n^{ab}n_{ab}=-2$. The
zeroth law of black-hole mechanics stating that $\kappa$ is constant
over the horizon \cite{Bardeen:1973gs,Racz:1995nh} is associated to
the Lorentz momentum map, just as the generalized zeroth law that
states that the electric potential is also constant over the horizon
in the Einstein-Maxwell theory is associated to the Maxwell momentum
map \cite{Elgood:2020svt}.\footnote{ This parallelism between zeroth
  laws was observed in \cite{Prabhu:2015vua}, also in the wider
  context of Einstein-Yang-Mills theories.} We are going to see that
further ``generalized zeroth laws'' are also associated to momentum
maps when we restrict ourselves to the bifurcation surface. We will
call them \textit{restricted generalized zeroth laws}.

Let us now consider the KR field. It is convenient to start by
considering the transformation of the 3-form field strength $H$
defined in Eq.~(\ref{eq:Hdef}) under diffeomorphisms. Since it is
gauge invariant, upon use of its Bianchi identity

\begin{equation}
    \delta_{\xi}H
    = 
    -\pounds_{\xi}H
    =
    -\imath_{\xi}dH -d\imath_{\xi}H
    =
    \imath_{\xi}\mathcal{F}_{I}\wedge\mathcal{F}^{I} -d\imath_{\xi}H\, .
\end{equation}

When $\xi=k$, this expression must vanish and we can use Eq.~(\ref{eq:Pkdef}),
which leads to the identity

\begin{equation}
\delta_{\xi}H
=
-d\left(\imath_{k}H +\mathcal{P}_{k\, I}\mathcal{F}^{I}
\right)
=
0\, ,
\end{equation}

\noindent
which, in turn, implies the local existence of a gauge-invariant 1-form
that we will also call a momentum map, satisfying

\begin{equation}
  \label{eq:KRPkdef}
  -\imath_{k}H -\mathcal{P}_{k\, I}\mathcal{F}^{I}
  =
  dP_{k}\, .
\end{equation}

The KR momentum map plays a fundamental role in the definition of the
variation of the KR 2-form $B$ under diffeomorphisms which should be of the
general form

\begin{equation}
  \delta_{\xi}B
  =
  -\pounds_{\xi}B +\left(\delta_{\Lambda_{\xi}}+\delta_{\chi_{\xi}} \right)B\, ,
\end{equation}

\noindent
where $\chi_{\xi}$ and $\Lambda_{\xi}$ are scalar and 1-form
parameters of compensating gauge transformations. They will
generically depend on $\mathcal{A}^{I}$ and $B$ as well as on
$\xi$. $\chi_{\xi}{}^{I}$ has to be the same parameter used in the
definition of the Lie-Maxwell derivative Eq.~(\ref{eq:LMparameter})
and we just have to determine $\Lambda_{\xi}$.  Now, the Maxwell and
Lorentz cases suggest that we try

\begin{equation}
  \Lambda_{\xi}
  =
  \imath_{\xi}B -P_{\xi}\, ,
\end{equation}

\noindent
which leads to 

\begin{equation}
  \begin{aligned}
    \delta_{\xi}B
    & =
    -\pounds_{\xi}B +d( \imath_{\xi}B -P_{\xi})
    +\tfrac{1}{2}\chi_{\xi\, I}d\mathcal{A}^{I}
    \\
    & \\
    & = -\left(\imath_{\xi}H+\mathcal{P}_{\xi\, I}\mathcal{F}^{I}+dP_{\xi}\right)
    +\tfrac{1}{2}\mathcal{A}_{I}\wedge \imath_{\xi}\mathcal{F}^{I}
    +\tfrac{1}{2}\mathcal{P}_{\xi\, I}\mathcal{F}^{I}\, .
  \end{aligned}
\end{equation}

When $\xi=k$, though,

\begin{equation}
  \begin{aligned}
    \delta_{k}B
    & =
    d\left(\tfrac{1}{2}\mathcal{P}_{k\, I} \mathcal{A}^{I}\right)\, .
  \end{aligned}
\end{equation}

\noindent
This is not zero but it can be absorbed into a redefinition of
$\Lambda_{\xi}$:

\begin{equation}\
  \label{eq:Lambdaxidef}
  \Lambda_{\xi}
  =
  \imath_{\xi}B -P_{\xi}-\tfrac{1}{2}\mathcal{P}_{k\, I} \mathcal{A}^{I}\, ,
\end{equation}

\noindent
which gives the variation

\begin{equation}
    \label{eq:variationB}
  \begin{aligned}
    \delta_{\xi}B 
    & =
    -\left(\imath_{\xi}H+\mathcal{P}_{\xi\, I}\mathcal{F}^{I}+dP_{\xi}\right)
    -\tfrac{1}{2}\mathcal{A}_{I}\wedge \delta_{\xi}\mathcal{A}^{I}\, .
  \end{aligned}
\end{equation}

\noindent
This form of the variation makes it evident that $\delta_{k}B=0$, because
$\delta_{k}\mathcal{A}^{I}=0$ and because of the definition of the KR momentum
map 1-form Eq.~(\ref{eq:KRPkdef}).

It remains to check that the vanishing of this variation is a gauge-invariant
statement. Indeed, if we perform a gauge transformation in $\delta_{\xi}B$,
taking into account that all the momentum maps and $\delta_{\xi}\mathcal{A}^{I}$
are gauge-invariant, we find

\begin{equation}
  \delta_{\rm gauge}\delta_{\xi}B
  =
  -\tfrac{1}{2}\delta_{\rm gauge}\mathcal{A}_{I}\wedge \delta_{\xi}\mathcal{A}^{I}\, ,
\end{equation}

\noindent
which vanishes identically for $\xi=k$. 

\subsection{The Wald-Noether charge}

The Wald-Noether charge is the conserved $(d-2)$-form associated to the
invariance of the action under diffeomorphisms
\cite{Wald:1993nt}. The transformations that we are going to consider
(combinations of standard Lie derivative and gauge transformations,
as we have explained) are

\begin{subequations}
  \label{eq:combinedtransformations}
  \begin{align}
    \delta_{\xi}\phi
    & =
      -\imath_{\xi}d\phi\, ,
      \\
    & \nonumber \\
        \delta_{\xi} \phi^{x}
     & =
     -\imath_{\xi}d\phi^{x}\, .
      \\
      & \nonumber \\
    \delta_{\xi}\mathcal{A}^{I}
    & =
      -\left(\imath_{\xi}\mathcal{F}^{I} +d\mathcal{P}_{\xi}{}^{I}\right)\, ,
    \\
    & \nonumber \\
    \delta_{\xi}e^{a}
    & =
      -\left(\mathcal{D}\xi^{a}+P_{\xi}{}^{a}{}_{b}e^{b}\right)\, ,
    \\
    & \nonumber \\
  \delta_{\xi}\omega^{ab}
     & =
      -\left(\imath_{\xi}R^{ab} +\mathcal{D}P_{\xi}{}^{ab}\right)\, ,
    \\
    & \nonumber \\
    \delta_{\xi}B+\tfrac{1}{2}\mathcal{A}_{I}\wedge \delta_{\xi}\mathcal{A}^{I}
    & =
    -\left(\imath_{\xi}H+\mathcal{P}_{\xi\, I}\mathcal{F}^{I}+dP_{\xi}\right)\, .
  \end{align}
\end{subequations}

From Eq.~(\ref{eq:deltaS}), and using the definition of
$\tilde{\mathbf{E}}_{I}$ in Eqs.~(\ref{eq:EI}) and (\ref{eq:tildeEI}) to
cancel the terms of the form
$\mathbf{E}_{B}\wedge\mathcal{A}_{I}\wedge\delta_{\xi}\mathcal{A}^{I}$, we get

\begin{equation}
  \label{eq:deltaxiS}
  \begin{aligned}
    \delta_{\xi} S & =
    -\int \left\{ \mathbf{E}_{a}\wedge 
      \left(\mathcal{D}\imath_{\xi}e^{a}+P_{\xi}{}^{a}{}_{b}e^{b}\right)
      +\mathbf{E}_{B}\wedge \left(\imath_{\xi}H
        +\mathcal{P}_{\xi\, I}\mathcal{F}^{I}+dP_{\xi}\right)
    \right.
    \\
    & \\
    &
    \hspace{1.3cm}
    \left.
      +\tilde{\mathbf{E}}_{I}\wedge
      \left(\imath_{\xi}\mathcal{F}^{I} +d\mathcal{P}_{\xi}{}^{I} \right)
      +\mathbf{E}_{\phi}\imath_{\xi}d\phi
      +\mathbf{E}_{x}\imath_{\xi}d \phi^{x}
    \right.
    \\
    & \\
    &
    \hspace{1.3cm}
    \left.
      -d\mathbf{\Theta}(\varphi,\delta_{\xi}\varphi) \right\}\, ,
\end{aligned}
\end{equation}

\noindent
while, from Eq.~(\ref{eq:Theta}), we get

 \begin{equation}
   \label{eq:Thetaxi}
  \begin{aligned}
    \mathbf{\Theta}(\varphi,\delta_{\xi}\varphi)
    & =
    e^{-2\phi}\star (e^{a}\wedge e^{b})\wedge
    \left(\imath_{\xi}R_{ab} +\mathcal{D}P_{\xi\, ab}\right)
    \\
    & \\
    &
    \hspace{.5cm}
    -2\imath_{a}de^{-2\phi}\star(e^{a}\wedge e^{b})\wedge
    \left(\mathcal{D}\xi_{b}+P_{\xi\, bc}e^{c}\right)
    \\
    & \\
    &
    \hspace{.5cm}
    +8e^{-2\phi}\star d\phi \imath_{\xi}d\phi
        -e^{-2\phi}g_{xy}\star d\phi^{y} \imath_{\xi}d\phi^{x}
    \\
    & \\
    &
    \hspace{.5cm}
    -e^{-2\phi}M_{IJ}\star\mathcal{F}^{J}\wedge
    \left(\imath_{\xi}\mathcal{F}^{I} +d\mathcal{P}_{\xi}{}^{I}\right)
    \\
    & \\
    &
    \hspace{.5cm}
    -e^{-2\phi}\star H\wedge
\left(\imath_{\xi}H+\mathcal{P}_{\xi\, I}\mathcal{F}^{I}+dP_{\xi}\right)\, .
  \end{aligned}
\end{equation}

Next, we consider the terms in $\delta_{\xi}S$ that contain momentum maps,
integrating by parts those which involve their derivatives:

\begin{equation}
  \begin{aligned}
    & \hspace{.5cm}
    \mathbf{E}_{a}\wedge P_{\xi}{}^{a}{}_{b}e^{b}
    +\tilde{\mathbf{E}}_{I}\wedge d\mathcal{P}_{\xi}{}^{I}
    +\mathbf{E}_{B}\wedge
    \left(\mathcal{P}_{\xi\, I}\mathcal{F}^{I}+dP_{\xi}\right)
    \\
    & \\
    & =
    \mathbf{E}^{[a}\wedge e^{b]} P_{\xi\, ab}
    +P_{\xi}d\mathbf{E}_{B}
    +(-1)^{d}\mathcal{P}_{\xi\, I}\left[d\tilde{\mathbf{E}}^{I}
      +(-1)^{d}\mathbf{E}_{B}\wedge \mathcal{F}^{I} \right]
    \\
    & \\
    & \hspace{.5cm}
    +d\left(P_{\xi}\wedge \mathbf{E}_{B}
      +(-1)^{d-1}\mathcal{P}_{\xi}{}^{I}\tilde{\mathbf{E}}_{I}\right)\, .
  \end{aligned}
\end{equation}

The terms in the first line vanish as a consequence of the Noether identities
Eqs.~(\ref{eq:noether1})-(\ref{eq:noether3}) and we are left with the total
derivative which will be added to
$\mathbf{\Theta}(\varphi,\delta_{\xi}\varphi)$. Thus, the variation of the
action takes the form

\begin{equation}
  \label{eq:deltaxiS2}
  \begin{aligned}
    \delta_{\xi} S & =
    -\int \left\{ \mathbf{E}_{a}\wedge\mathcal{D}\imath_{\xi}e^{a}
      +\mathbf{E}_{B}\wedge\imath_{\xi}H
     +\tilde{\mathbf{E}}_{I}\wedge\imath_{\xi}\mathcal{F}^{I}
      +\mathbf{E}_{\phi}\imath_{\xi}d\phi
      +\mathbf{E}_{x}\imath_{\xi}d \phi^{x}
    \right.
    \\
    & \\
    &
    \hspace{1.3cm}
    \left.
      -d\left[\mathbf{\Theta}(\varphi,\delta_{\xi}\varphi)
        -P_{\xi}\wedge \mathbf{E}_{B}
      +(-1)^{d}\mathcal{P}_{\xi}{}^{I}\tilde{\mathbf{E}}_{I}
      \right]\right\}\, .
\end{aligned}
\end{equation}

\noindent
Integrating the first term of Eq.~(\ref{eq:deltaxiS2}) by parts we get another
total derivative to add to $\mathbf{\Theta}(\varphi,\delta_{\xi}\varphi)$ and
($\imath_{\xi}e^{a}=\xi^{a}$)

\begin{equation}
(-1)^{d}\mathcal{D}\mathbf{E}_{a}\xi^{a}
      +\mathbf{E}_{B}\wedge\imath_{\xi}H
     +\tilde{\mathbf{E}}_{I}\wedge\imath_{\xi}\mathcal{F}^{I}
     +\mathbf{E}_{\phi}\imath_{\xi}d\phi
     +\mathbf{E}_{x}\imath_{\xi}d \phi^{x}=0\, ,  
\end{equation}

\noindent
by virtue of the Noether identity associated to the invariance under
diffeomorphisms and, therefore,

\begin{equation}
  \label{eq:deltaxiS3}
  \begin{aligned}
    \delta_{\xi} S & =
    \int d\mathbf{\Theta}'(\varphi,\delta_{\xi}\varphi)\, ,
\end{aligned}
\end{equation}

\noindent
where

\begin{equation}
  \label{eq:Thetaprime}
  \mathbf{\Theta}'(\varphi,\delta_{\xi}\varphi)
  =
  \mathbf{\Theta}(\varphi,\delta_{\xi}\varphi)
  +(-1)^{d}\mathbf{E}_{a}\xi^{a}
  -P_{\xi}\wedge \mathbf{E}_{B}
      +(-1)^{d}\mathcal{P}_{\xi}{}^{I}\tilde{\mathbf{E}}_{I}\, .
\end{equation}

Usually, the last three terms, which are proportional to equations of motion
and vanish on-shell, are ignored for this very reason. However, we have found
that keeping them is actually quite useful for finding the Wald-Noether charge,
because they are exactly what is needed to write $\mathbf{J}$ as a total
derivative. Without them, we would have had to guess which combinations of the
equations of motion should be added to achieve that goal. Furthermore, the
result that we will obtain will be valid off-shell.

Since the action is exactly invariant under the gauge transformations
Eq.~(\ref{eq:gaugetransformations}), but it is only invariant up to a total
derivative under standard infinitesimal diffeomorphisms, under the
combined transformations Eqs.~(\ref{eq:combinedtransformations})

\begin{equation}
\delta_{\xi}S
=
-\int d\imath_{\xi}\mathbf{L}\, ,
\end{equation}

\noindent
which, combined with Eq.~(\ref{eq:deltaxiS3}), leads to the identity

\begin{equation}
  \label{eq:dJ=0}
  d\mathbf{J}=0\, ,
\end{equation}

\noindent
which holds off-shell for arbitrary $\xi$ with

\begin{equation}
  \label{eqLJ0def}
  \mathbf{J}
  \equiv
  \mathbf{\Theta}'(\varphi,\delta_{\xi}\varphi)
  +\imath_{\xi}\mathbf{L}\, .  
\end{equation}

\noindent
Eq.~(\ref{eq:dJ=0}) implies the local existence of a $(d-2)$-form
$\mathbf{Q}[\xi]$ such that

\begin{equation}
  \label{eq:J=dQ}
  \mathbf{J}
  =
  d\mathbf{Q}[\xi]\, .
\end{equation}

Using the previous results we find that, up to total derivatives and up to the
overall factor $(g^{(d)\, 2}_{s}16\pi G^{(d)}_{N})^{-1}$ that we are suppressing to get
simpler expressions

 \begin{equation}
   \label{eq:Q}
  \begin{aligned}
    \mathbf{Q}[\xi]
        & =
        (-1)^{d}\star (e^{a}\wedge e^{b})
        \left[e^{-2\phi}P_{\xi\, ab}-2\imath_{a}de^{-2\phi}\xi_{b}\right]
    \\
    & \\
    & \hspace{.5cm}
      +(-1)^{d-1}\mathcal{P}_{\xi}{}^{I}
      \left(e^{-2\phi}M_{IJ}\star\mathcal{F}^{J}\right)
      -P_{\xi}\wedge \left(e^{-2\phi}\star H\right)\, .
  \end{aligned}
\end{equation}

\section{Zeroth laws}
\label{sec-zeroth}

The zeroth law and its generalizations, ensuring that the surface gravity and
the electrostatic potential are constant over the event (Killing) horizon
$\mathcal{H}$ are important ingredients in the standard derivation of the
first law of black-hole mechanics in the context of the Einstein-Maxwell
theory \cite{Bardeen:1973gs}. In presence of higher-rank $p$-form fields, it
is not clear how these laws should be further generalized. However, it is
possible to proof the first law using Wald's formalism working on the
bifurcation sphere $\mathcal{BH}$, where the Killing vector $k$ associated to
the horizon vanishes. This restricts the validity of the proof to bifurcate
horizons but, on the other hand, it makes it possible to carry out the proof
using a more restricted form of the (generalized) zeroth laws which states the
closedness of the electrostatic potential and its higher-rank generalizations
on $\mathcal{BH}$. Since the electrostatic potential is a scalar, its
closedness implies that it is constant on $\mathcal{BH}$, which is a
restricted version of the generalized zeroth law. For higher-rank potentials
closedness is, actually, all we need, as we will see in the next section.

We start by assuming that all the field strengths of the theory are regular on
the horizon.\footnote{Observe that in this theory in which all the field
  strengths are gauge-invariant, this is a gauge-invariant statement that
  should be valid in a regular coordinate patch.} This implies that

\begin{subequations}
  \begin{align}
  \imath_{k}\mathcal{F}^{I}
    & \stackrel{\mathcal{BH}}{=} 0\, ,
  \\
    & \nonumber \\
    \label{eq:ikH2}
  \imath_{k}H
    & \stackrel{\mathcal{BH}}{=} 0\, .
  \end{align}
\end{subequations}

\noindent
The first equation directly implies the closedness of the components of the
momentum map $\mathcal{P}^{I}_{k}$ on $\mathcal{BH}$ on account of its
definition Eq.~(\ref{eq:Pkdef}), and, hence, its constancy on $\mathcal{BH}$,
a statement that we can call \textit{restricted generalized zeroth law} after
the natural identification of $\mathcal{P}^{I}_{k}$ with the electrostatic
black-hole potential $\Phi^{I}$.  Observe that, our gauge-invariant definition
of the electrostatic black-hole potential guarantees that it is fully defined
up to an additive constant that can be determined by setting the value of the
potential at infinity to zero.

Using Eq.~(\ref{eq:ikH2}) and the constancy of $\mathcal{P}^{I}_{k}$ on on
$\mathcal{BH}$ in the definition of the KR momentum map Eq.~(\ref{eq:KRPkdef})
we find that 

\begin{equation}
  0 \stackrel{\mathcal{BH}}{=} -\imath_{k}H
  = dP_{k} +\mathcal{P}_{k\, I}\mathcal{F}^{I}
  \stackrel{\mathcal{H}}{=}
  d\left(P_{k} +\mathcal{P}_{k\, I}\mathcal{A}^{I}\right)\, .
\end{equation}

We can call the combination $P_{k} +\mathcal{P}_{k\, I}\mathcal{A}^{I}$ that
is closed on $\mathcal{BH}$ the KR black-hole potential $\Phi$ and its
closedness can be understood as another restricted generalized zeroth law of
black-hole mechanics in this theory. Observe that $\Phi$ is not
gauge-invariant, but $P_{k}$ is only defined up to shifts by exact 1-forms
anyway and, when we use $\Phi$ as the 1-form $\Lambda$ in the calculation of
the KR charge Eq.~(\ref{eq:QLambdacharge-0}), the addition of exact 1-forms
does not change the value of the associated KR charge
Eq.~(\ref{eq:QLambdacharge}). The fact that this $\Phi$ occurs in the
expressions leading to the first law precisely plays this role is quite a
non-trivial check of the consistency of our results.

\section{The first law}
\label{sec-firstlaw}

We start by defining the \textit{pre-symplectic $(d-1)$-form}
\cite{Lee:1990nz}

\begin{equation}
\omega(\varphi,\delta_{1}\varphi,\delta_{2} \varphi)  
\equiv 
\delta_{1}\mathbf{\Theta}(\varphi,\delta_{2} \varphi)
-\delta_{2}\mathbf{\Theta}(\varphi,\delta_{1} \varphi)\, ,
\end{equation}

\noindent
and the \textit{symplectic form} relative to the Cauchy surface $\Sigma$

\begin{equation}
\Omega(\varphi,\delta_{1}\varphi,\delta_{2} \varphi)  
\equiv
\int_{\Sigma}\omega(\varphi,\delta_{1}\varphi,\delta_{2} \varphi)\, .
\end{equation}

Now, following Ref.~\cite{Iyer:1994ys}, when $\varphi$ solves the equations of
motion $\mathbf{E}_{\varphi}=0$ if $\delta_{1}\varphi=\delta\varphi$ is an
arbitrary variation of the fields and $\delta_{2}\varphi= \delta_{\xi}\varphi$
is their variation under diffeomorphisms, we have that

\begin{equation}
  \omega(\varphi,\delta\varphi,\delta_{\xi}\varphi)
  =
  \delta\mathbf{J}+d\imath_{\xi}\mathbf{\Theta}'
  =
  \delta d\mathbf{Q}[\xi]+d\imath_{\xi}\mathbf{\Theta}'\, ,
\end{equation}

\noindent
where, in our case, $\mathbf{J}=d\mathbf{Q}$, where $\mathbf{Q}$ is given by
Eq.~(\ref{eq:Q}) and $\mathbf{\Theta}'$ is given in
Eq.~(\ref{eq:Thetaprime}). Since, on-shell,
$\mathbf{\Theta} = \mathbf{\Theta}'$, we have that, if $\delta\varphi$
satisfies the linearized equations of motion,
$\delta d\mathbf{Q}= d\delta \mathbf{Q}$.  Furthermore, if the parameter
$\xi=k$ generates a transformation that leaves invariant the field
configuration, $\delta_{k}\varphi=0$,\footnote{We have constructed variations
  of the fields $\delta_{\xi}$ for which this is possible.} linearity implies
that $\omega(\varphi,\delta\varphi,\delta_{k}\varphi)=0$, and

\begin{equation}
d\left( \delta \mathbf{Q}[k]+\imath_{k}\mathbf{\Theta}'  \right)=0\, .
\end{equation}

\noindent
Integrating this expression over a hypersurface $\Sigma$ with boundary
$\delta\Sigma$ and using Stokes' theorem we arrive at

\begin{equation}
  \int_{\delta\Sigma}
  \left( \delta \mathbf{Q}[k]+\imath_{k}\mathbf{\Theta}'  \right)
  =
  0\, .
\end{equation}

We are interested in asymptotically flat, stationary, black-hole spacetimes
and we choose $k$ as the Killing vector whose Killing horizon coincides with
the event horizon $\mathcal{H}$, which we assume to be a bifurcate
horizon. This Killing vector $k$ is assumed to be linear combination with
constant coefficients $\Omega^{n}$ of the timelike Killing vector associated to
stationarity, $t^{\mu}\partial_{\mu}$ and the $[\tfrac{1}{2}(d-1)]$
inequivalent rotations $\phi_{n}^{\mu}\partial_{\mu}$

\begin{equation}
  k^{\mu} = t^{\mu} +\Omega^{n}\phi_{n}^{\mu}\, .  
\end{equation}

\noindent
Furthermore, we choose the hypersurface $\Sigma$ to be the space between
infinity and the bifurcation sphere ($\mathcal{BH}$) on which $k=0$. Then, its
boundary $\delta\Sigma$ has two disconnected pieces: a $(d-2)$-sphere at
infinity, S$^{d-2}_{\infty}$, and the bifurcation sphere $\mathcal{BH}$.
Then, taking into account that $k=0$ on $\mathcal{BH}$, we obtain the relation

\begin{equation}
\delta \int_{\mathcal{BH}}   \mathbf{Q}[k]
=
\int_{\mathrm{S}^{d-2}_{\infty}}
  \left( \delta \mathbf{Q}[k]+\imath_{k}\mathbf{\Theta}'  \right)\, .
\end{equation}

As explained in Ref.~\cite{Iyer:1994ys,Compere:2007vx}, the right-hand side
can be identified with $\delta M -\Omega^{m}\delta J_{n}$, where $M$ is the
total mass of the black-hole spacetime and $J_{n}$ are the independent
components of the angular momentum.

Using the explicit form of $\mathbf{Q}[k]$, Eq.~(\ref{eq:Q}), and restoring
the overall factor $g^{(d)\, 2}_{s}(16\pi G^{(d)}_{N})^{-1}$, we find

\begin{equation}
  \begin{aligned}
    \delta \int_{\mathcal{BH}} \mathbf{Q}[k]
    & =
    \frac{(-1)^{d-1}g^{(d)\, 2}_{s}}{16\pi
      G^{(d)}_{N}}\delta \int_{\mathcal{BH}}
    \mathcal{P}_{k}{}^{I}\left(e^{-2\phi}M_{IJ}\star\mathcal{F}^{J}\right)
    \\
    & \\
    & \hspace{.5cm}
    -\frac{g^{(d)\, 2}_{s}}{16\pi G^{(d)}_{N}}\delta \int_{\mathcal{BH}} P_{k}\wedge
    \left(e^{-2\phi}\star H\right)
    \\
    & \\
    & \hspace{.5cm}
    +\frac{(-1)^{d}g^{(d)\, 2}_{s}}{16\pi G^{(d)}_{N}} \delta
    \int_{\mathcal{BH}} \star (e^{a}\wedge e^{b})
        \left[e^{-2\phi}P_{k\, ab}-2\imath_{a}de^{-2\phi}k_{b}\right]\, .
  \end{aligned}
\end{equation}

The last term vanishes over the bifurcation sphere and will be removed
from now on.

As it is, this expression has two problems that make it difficult for us to
obtain the kind of terms that occur in the first law. In the first line, we
have an expression that we should be able to interpret in terms of the
electric charges $\mathcal{Q}_{I}$. However, when we compare this with
Eq.~(\ref{eq:QIdef}) we see that the second term in the integrand is
missing. Without that term, the charge is not conserved. On the other hand, in
the second line, we have an expression that we should be able to interpret in
terms of the KR charge using Eq.~(\ref{eq:QLambdacharge-0}). However, the
1-form $P_{k}$ is not closed on $\mathcal{BH}$.

The solution to these two problems is unique: the addition and
subtraction of the term
$\mathcal{P}_{k\, I}\mathcal{A}^{I}\wedge \left(e^{-2\phi}\star
  H\right)$ in the integrand, so that the integral to evaluate on
$\mathcal{BH}$ takes the form

\begin{equation}
  \begin{aligned}
    \delta \int_{\mathcal{BH}} \mathbf{Q}[k]
    & =
    \frac{(-1)^{d-1}g^{(d)\, 2}_{s}}{16\pi
      G^{(d)}_{N}}\delta \int_{\mathcal{BH}}
    \mathcal{P}_{k}{}^{I}\left[e^{-2\phi}M_{IJ}\star\mathcal{F}^{J}
   +e^{-2\phi}\star H\wedge\mathcal{A}_{I}
    \right]
    \\
    & \\
    & \hspace{.5cm}
    -\frac{g^{(d)\, 2}_{s}}{16\pi G^{(d)}_{N}}\delta \int_{\mathcal{BH}}
    \left(P_{k}+\mathcal{P}_{k\, I}\mathcal{A}^{I}\right)\wedge
    \left(e^{-2\phi}\star H\right)
    \\
    & \\
    & \hspace{.5cm}
    +\frac{(-1)^{d}g^{(d)\, 2}_{s}}{16\pi G^{(d)}_{N}} \delta
    \int_{\mathcal{BH}} e^{-2\phi}\star (e^{a}\wedge e^{b})P_{k\, ab}\, .
  \end{aligned}
\end{equation}

Now, using the generalized zeroth law that ensures that
$\mathcal{P}_{k}{}^{I}\equiv \Phi^{I}$ is constant over $\mathcal{H}$,
in particular on $\mathcal{BH}$, and the definition of electric
charge Eq.~(\ref{eq:QIdef}), the first term in the right-hand side
takes the form

\begin{equation}
\Phi^{I}\delta \mathcal{Q}_{I}\, .
\end{equation}

\noindent
Next, from the closedness of the combination
$\Phi=P_{k}+\mathcal{P}_{k\, I}\mathcal{A}^{I}$ on $\mathcal{BH}$, (the
restricted generalized zeroth law) using the Hodge decomposition

\begin{equation}
  P_{k}+\mathcal{P}_{k\, I}\mathcal{A}^{I} \stackrel{\mathcal{BH}}{=}
  d e + \Phi^{i}\Lambda_{h\, i}\, ,
\end{equation}

\noindent
where the $\Lambda_{h\, i}$ are harmonic 1-forms on $\mathcal{BH}$
and the $\Phi^{1}$ are constants that have the interpretation of
potentials associated to the charge of the KR field (the dipole charge
of Ref.~\cite{Emparan:2004wy} in particular), and using the definition
Eq.~(\ref{eq:QLambdacharge}), we find that the second term in the right-hand side
takes the form

\begin{equation}
  \Phi^{i}\delta Q_{i}\, ,
  \hspace{1cm}
   Q_{i} \equiv Q[\Lambda_{h\, i}]\, .
\end{equation}

Observe that the addition and subtraction of the term
$\mathcal{P}_{k\, I}\mathcal{A}^{I}\wedge \left(e^{-2\phi}\star
  H\right)$ has been crucial to recover the correct definition of the
charges which, in particular, demands the occurrence of the closed
1-form $P_{k}+\mathcal{P}_{k\, I}\mathcal{A}^{I}$.

Now, let us consider the third integral. Before we compute it
explicitly, we notice that the integrand is identical, up to a sign,
to the Lorentz charge Eq.~(\ref{eq:Lorentzconservedd-2form}) computed
for the Lorentz parameter $P_{k}{}^{a}{}_{b}$ which is covariantly
constant over the bifurcation surface. This coincidence is very
intriguing and will be further explored in Ref.~\cite{Elgood:2020nls}.

Using Eq.~(\ref{eq:Pkab})

\begin{equation}
  \label{eq:TdA}
  \begin{aligned}
    \frac{(-1)^{d}\kappa}{16\pi G^{(d)}_{N}}
    \delta\int_{\mathcal{BH}}e^{-2(\phi-\phi_{\infty})}\star
    (e^{a}\wedge e^{b})n_{ab} =\,\, & -\frac{\kappa}{16\pi
      G^{(d)}_{N}}
    \delta\int_{\mathcal{BH}}e^{-2(\phi-\phi_{\infty})}n^{ab}n_{ab}
    \\
    & \\
    =\,\, & T \delta \frac{\mathcal{A}_{\mathcal{H}}}{4 G^{(d)}_{N}}\, ,
  \end{aligned}
\end{equation}

\noindent
where we have used the normalization of the binormal
$n_{ab}n^{ab}=-2$, $T=\kappa/2\pi$ is the Hawking temperature and

\begin{equation}
  \mathcal{A}_{\mathcal{H}}
  \equiv
  \int_{\mathcal{B}}d^{d-2}S e^{-2(\phi-\phi_{\infty})}\, ,
\end{equation}

\noindent
is the area of the horizon measured with the \textit{modified Einstein
  frame metric} \cite{Maldacena:1996ky} which is obtained from the
string one by multiplying by the conformal factor
$e^{-4(\phi-\phi_{\infty})/(d-2)}$, and computed using the spatial
section $\mathcal{BH}$.

We finally get the following expression for the first law of black
hole mechanics in the Heterotic Superstring effective action to
leading order in $\alpha'$:

\begin{equation}
\delta M
    =
    T\delta  \frac{\mathcal{A}_{\mathcal{H}}}{4G_{N}^{(d)}}
    +\Omega^{m}\delta J_{m}
    +\Phi^{i}\delta Q_{i}
    +\Phi^{I}\delta \mathcal{Q}_{I}\, ,
\end{equation}

\noindent
which leads to the interpretation of the area of the horizon divided by
$4G_{N}^{(d)}$ as the black-hole entropy.

\section{Momentum Maps for Black Rings in $d=5$}
\label{sec-example}

In this section we are goin to illustrate how the definitions made and the
properties proven in the previous sections work in an explicit example. In
particular, we are going to determine the values of the momentum maps,
checking the restricted generalized zeroth laws.

The solution we are going to consider is a non-extremal, charged, black ring
solution of pure $\mathcal{N}=1,d=5$ supergravity which can be easily embedded
in the toroidally-compactified Heterotic Superstring effective field theory
using the results in Appendix~\ref{sectruncationtoN1d5SUGRA}. This embedding
is necessary because all the definitions and formulae that we have developed
are adapted to that theory. In Appendix~\ref{sectruncationtoN1d5SUGRA} we show
how the action Eq.~(\ref{eq:heterotic(10-n)order0}), for $d=5$ can be
consistently truncated to that of pure $\mathcal{N}=1,d=5$ supergravity
Eq.~\ref{eq:pureN1d5action} in two steps:

\begin{enumerate}
\item A direct truncation of some fields of the Heterotic theory, to obtain a
  model of $\mathcal{N}=1,d=5$ supergravity coupled to two vector
  multiplets. The Kalb-Ramond 2-form has to be dualized into a 1-form in order
  to obtain the supergravity theory in the standard form, with 3 1-forms which
  can be treated on the same footing and which may be linearly combined.
  
\item A consistent truncation of the two vector supermultiplets. In this
  truncation, rather than setting two of the vector fields to zero, they are
  identified with the surviving vector, up to numerical factors. This allows
  the scalars in the vecort supermultiplets to take their vacuum values.
\end{enumerate}

Given a solution of pure $\mathcal{N}=1,d=5$ supergravity, one can easily
retrace those steps, restoring, first, the two ``matter'' vector fields so the
solution becomes now a solution of $\mathcal{N}=1,d=5$ supergravity coupled to
two vector multiplets. Then, dualizing the vector in the supergravity
multiplet to recover the Kalb-Ramond 2-form, the solution can immediately be
interpreted as a solution of the Heterotic Superstring effective field theory
in which many other fields simply take their vacuum values.

The non-extremal, charged, black ring solution that we are going to consider
is the one given in Section~4 of Ref.~\cite{Elvang:2004xi}.  This solution
belongs to a more general family of non-supersymmetric black rings with three
charges $\alpha_{i}$, three dipoles $\mu_{i}$, with $i=1,2,3$, and two angular
momenta $J_{\varphi}$ and $J_{\psi}$ in the theory with two vector
supermultiplets. The solution above corresponds to setting all three charges
and three dipoles equal, $\alpha_{i}=\alpha$ and $\mu_{i}=\mu$ for all
$i$. This identification of the charges and dipoles coprresponds to the
identification between the vector fields that leads from the supergravity
theory with matter to the theory of pure supergravity.  Let us review the
solution and its main features.

The physical fields of the solution (the metric and the Abelian connection
$A$) can be written in terms of the five parameters $(R,\alpha,\mu,\lambda,\nu)$
(all of them dimensionless except for the length scale $R$) and the three
functions, $F(\xi),H(\xi)$ and $G(\xi)$, given by

\begin{equation}
  \label{eq:fundamentalfunctions}
  H(\xi) = 1-\mu\xi\, ,
  \hspace{1cm}
  F(\xi) = 1+\lambda\xi\, ,
  \hspace{1cm}
  G(\xi) = (1-\xi^{2}) (1+\nu\xi)\, .
\end{equation}

The line element is

\begin{align}
  \label{lineelement}
    ds^{2}
    & =
      \frac{U(x,y)}{h_{\alpha}^{2}(x,y)}
      \left(dt +\omega_{\psi}(y)d\psi +\omega_{\varphi}(x)d\varphi\right)^{2}
      -h_{\alpha}(x,y) F(x) H(x) H(y)^{2} \times
      \nonumber \\
    & \nonumber \\
    & \hspace{.5cm}
      \times
       \frac{R^{2}}{(x-y)^{2}}\left[
      -\frac{G(y)}{F(y)H(y)^{3}}d\psi^{2} -\frac{dy^{2}}{G(y)}
      +\frac{dx^{2}}{G(x)}
      +\frac{G(x)}{F(x)H(x)^{3}}d\varphi^{2}
      \right]\, ,
\end{align}

\noindent
where we use the shorthand notation $s=\sinh{\alpha}$ and $c=\cosh{\alpha}$,
the following combinations of the fundamental parameters 

\begin{equation}
  C_{\lambda}
  =
  \epsilon_{\lambda}\sqrt{\lambda(\lambda-\nu)\frac{1+\lambda}{1-\lambda}}\,,
\,\,\,\,
C_{\mu}
=
\epsilon_{\mu}\sqrt{\mu(\mu+\nu)\frac{1-\mu}{1+\mu}}\,,
\,\,\,\,
\epsilon_{\lambda,\mu}
=
\pm 1\,,
\end{equation}

\noindent
and the following combinations of the fundamental functions in
Eq.~(\ref{eq:fundamentalfunctions})

\begin{subequations}
  \begin{align}
    U(x,y)
    & =
      \frac{H(x)}{H(y)}\frac{F(y)}{F(x)}\, ,
    \\
    & \nonumber \\
    h_{\alpha}(x,y)
    & =
      1 +\frac{(\lambda+\mu)(x-y)}{F(x)H(y)}s^{2}\, ,
    \\
    & \nonumber \\
    \omega_{\psi}(y)
    & =
      R(1+y) \left[\frac{1}{F(y)}C_{\lambda}c^{3}-
      \frac{3}{H(y)}C_{\mu}cs^{2}\right]\, ,
    \\
    & \nonumber \\
    \omega_{\varphi}(x)
    & =
    -R (1+x) s\left(\frac{1}{F(x)}C_{\lambda} s^2-\frac{3}{H(x)} C_{\mu}c^2
      \right)\, .
  \end{align}
\end{subequations}

Finally, the gauge field reads

\begin{subequations}
  \begin{align}
    -A/\sqrt{3}
    & =
      \frac{U(x,y)-1}{h_{\alpha}(x,y)}cs dt
      \nonumber \\
    & \nonumber \\
    & \hspace{.5cm}
      +\frac{R(1+y)}{h_{\alpha}(x,y)}
      \left[
      \frac{U(x,y)}{F(y)}C_{\lambda}c^{2}s
      -\frac{U(x,y)}{H(y)}C_{\mu}s^{3}
      -\frac{2}{H(y)}C_{\mu}c^{2}s
      \right]d\psi
      \nonumber \\
    & \nonumber \\
    & \hspace{.5cm}
      +\frac{R(1+x)}{h_{\alpha}(x,y)}
      \left[
      2\frac{U(x,y)}{H(x)}C_{\mu}cs^{2}
      -\frac{1}{F(x)}C_{\lambda}cs^{2}
      +\frac{1}{H(x)}C_{\lambda}c^{3}
      \right]d\varphi\, .
  \end{align}
\end{subequations}

The parameters of the solution must satisfy the constraints

\begin{equation}
  0<\nu\leq\lambda<1\,,
    \hspace{1.5cm}
0\leq\mu<1\,,
\end{equation} 

\noindent
to avoid naked singularities. Additional constraints arise from the codition
of absence of Dirac-Misner strings and conical sigularities, as we are going
to see.

The coordinates $x,y$ take values in

\begin{equation}
  -\infty<y\leq-1\,,
  \hspace{1.5cm}
  -1\leq x\leq 1\,.
\end{equation}

\begin{figure}[h!]
\centering
\includegraphics[scale=0.4]{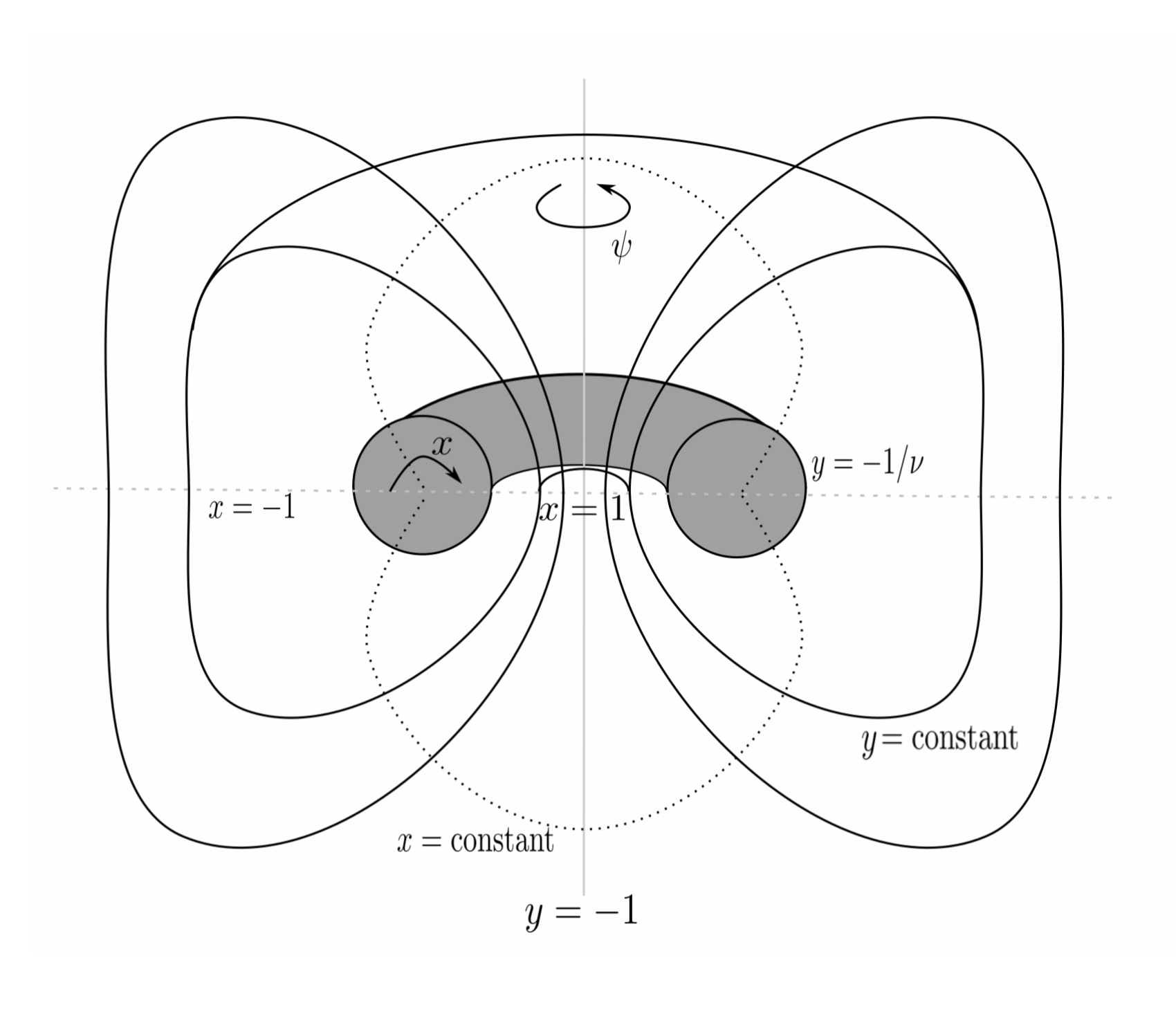}
\caption{Sketch of a section of constant $t$ and $\varphi$ of the black ring
  (figure based on Ref.~\cite{Emparan:2004wy}). The disc at $x=1$ and infinite
  annulus at $x=-1$ are the axes (fixed points) of $\partial_{\varphi}$, while
  the axis of the ring is at $y=-1$ (fixed points of
  $\partial_{\psi}$). Surfaces of constant $y$ have topology
  $S^{1}\times S^{2}$. $y=-1/\nu$ corresponds to the horizon (shaded surface)
  while surfaces of constant $y\in(-1/\nu,-1)$ are fatter rings containing the
  horizon in their interior.}
\label{fig1}
\end{figure}

\noindent
The surfaces of constant $y$ have the topology S$^{2}\times$S$^{1}$. $x$ is a
polar coordinate on the S$^{2}$ (essentially, $x\sim \cos{\theta}$), which is
also parametrized by $\varphi$, which plays the role of azymuthal angle.  $\psi$
parametrizes the S$^{1}$, see Fig.~\ref{fig1}.  Spatial infinity is approached
when both $x$ and $y$ go to $-1$, although the coordinates are ill-defined in
that limit.\footnote{Good coordinates at infinity can be found in
  Ref.~\cite{Elvang:2004xi}.} The orbits of the vector $\partial_{\varphi}$ close
off at $x=-1$, but do not do the same at $x=1$ unless $\omega_{\varphi}(x=+1)=0$,
which can forces us to require

\begin{equation}
  \label{DMstrings}
\frac{C_{\lambda}}{1+\lambda}s^{2} = \frac{3C_{\mu}}{1-\mu}c^{2}\,,
\end{equation}

\noindent
which removes any possible Dirac-Misner strings. (The same constraint makes
$A_{\varphi}(x=+1)$ independent of $y$.) Then, the fixed point sets of
$\partial_{\psi}$ and $\partial_{\varphi}$ are, respectively, $y=-1$ (axis of the
ring) and $x=1,-1$ (inner and outer axes of the S$^{2}$).

Finally, the periods of $\psi$ and $\varphi$ must be chosen appropriately so as
to avoid conical singularities. The axes $y=-1$ and $x=-1$ (which extend to
infinity) are regular for the periods

\begin{equation}
  \label{periods}
\Delta\psi=\Delta\varphi=2\pi\frac{\sqrt{1-\lambda}}{1-\nu}\left(1+\mu\right)^{3/2}\,.
\end{equation}

For generic values of the parameters, though, the period of $\varphi$ required
by smoothness at the inner axis, $x=1$, differs from the above
$\Delta\varphi$. Making both periods coincide (``balancing'' the ring) is
possible only when the following constraint holds

\begin{equation}
  \label{Balance}
  \left(\frac{1-\nu}{1+\nu}\right)^{2}
  =
  \frac{1-\lambda}{1+\lambda}\left(\frac{1+\mu}{1-\mu}\right)^{3}\,.
\end{equation}

Henceforth we shall assume that Eqs.~(\ref{DMstrings}) and (\ref{Balance})
hold, so that, effectively, we will be dealing with a three-parameter family
of solutions. As shown in Ref.~\cite{Elvang:2004xi}, the mass, the two
independent angular momenta and the area of the event horizon of the solution
read

\begin{subequations}
  \begin{align}
    M
    & =
      \frac{3\pi R^{2}}{4G_{N}^{(5)}}
      \frac{(\lambda+\mu)(1+\mu)^{2}}{1-\nu}\cosh{2\alpha}\, ,
    \\
    & \nonumber \\
    J_{\psi} 
    & =
      \frac{\pi R^{3}}{2G_{N}^{(5)}}
\frac{(1-\lambda)^{3/2}(1+\mu)^{9/2}}{(1-\nu)^{2}}
\left[  \frac{C_{\lambda}}{1-\lambda} c^{3}-\frac{3C_\mu}{1+\mu}s^{2}c \right]\, ,
    \\
    & \nonumber \\
    J_{\varphi}
    & =
      -\frac{3\pi R^{3}}{G_{N}^{(5)}}
\frac{\sqrt{1-\lambda}\;(1+\mu)^{7/2}(\lambda+\mu)}{(1-\nu)^{2}(1-\mu)}
      C_{\mu}c^{2}s\, ,
    \\
    & \nonumber \\
    \mathcal{A}_{\mathcal{H}}
    & = 8 \pi^{2} R^{3} \,
  \frac{(1-\lambda)(\lambda - \nu)^{1/2}(1+\mu)^{3} (\nu+ \mu)^{3/2} }{(1-\nu)^{2}(1+\nu)}
 \left|
  \frac{C_{\lambda}}{\lambda-\nu} c^{3}
  +\frac{3C_{\mu}}{\nu+\mu} s^{2}c
      \right|\,.
  \end{align}
\end{subequations}

There is an ergosurface at $y=-1/\lambda$, where the norm of $\partial_{t}$
vanishes, and the event horizon lies at $y=-1/\nu$. It is a Killing horizon of

\begin{equation}
  \label{eq:kvectorBR}
k=\partial_{t}+\Omega\partial_{\psi},
\end{equation}

\noindent
where $\Omega$, the angular velocity of the horizon in the direction $\psi$,
can be conveniently written as
$\Omega = -1/\omega_{\psi}(-1/\nu)$.\footnote{Notice we work with coordinates
  $\varphi,\psi$ whose periods are not the standard ones, but those given in
  Eq.~(\ref{periods}).} A rather unusual property of this solution is that
the horizon has no angular velocity in the direction $\varphi$ even though
$J_{\varphi}\neq 0$. Finally, the horizon temperature is

\begin{equation}
    T_{\mathcal{H}}^{-1}
     =
      4\pi R\frac{\sqrt{\lambda-\nu}(\mu+\nu)^{3/2}}{\nu(1+\nu)}
\left|
  \frac{C_\lambda}{\lambda-\nu} c^{3}
  +\frac{3C_\mu}{\nu+\mu} s^{2}c
      \right|\, .
\end{equation}

This solution of pure $\mathcal{N}=1,d=5$ supergravity corresponds to a
following solution of the Heterotic Superstring effective field theory
compactified on T$^{4}\times$S$^{1}$ with the same metric and the non-trivial
matter fields given by\footnote{The fields that arise in the compactification
  over T$^{4}$ and which are set to their vacuum values (they are trivial)
  have not been considered. In particular, the index $I$ takes only two values
  because the fields corresponding to the other values are trivial.}

\begin{subequations}
  \begin{align}
    \phi
    & =
      \phi_{\infty}\,,
    \\
    & \nonumber \\
    M_{IJ}
    & =
      \left(\begin{matrix}k_{\infty}^{2} & 0\\ 0& k_{\infty}^{-2}\end{matrix}\right)\,,
    \\
    & \nonumber \\
    \mathcal{A}^{I}
    & =\left(\begin{matrix}k_{\infty}^{-1}\\ k_{\infty} \end{matrix}\right) \mathcal{A}\,,    
    \\
    & \nonumber \\
    H
    & =  d B-\tfrac{1}{2}\mathcal{A}_{I}\wedge\mathcal{F}^{I}=\star \mathcal{F}  
  \end{align}
\end{subequations}

\noindent
where, for convenience, we have introduced $\mathcal{A}=-A/\sqrt{3}$ and its
field strength $ \mathcal{F} = d \mathcal{A}$. Let us obtain the vector and KR
momentum maps asociated to the Killing vector $k$ in Eq.~(\ref{eq:kvectorBR})
for this solution, denoted, respectively, as $\mathcal{P}_{k}{}^{I}$ and
$P_{k}$. In the following we consider a constant $t$ surface $\Sigma$ defined
by which extends from the bifurcate surface (here, a ring) $\mathcal{BH}$ at
$y=-1/\nu$ to infinity (analogously to one leaf of the Einstein--Rosen
bridge). The vector momentum maps $\mathcal{P}_{k}^{I}$ can be written as

\begin{equation}\label{PtoPI}
  \mathcal{P}_{k}^{I}
  =
  \left(\begin{matrix}k_{\infty}^{-1}\\k_{\infty} \end{matrix}\right)
  \mathcal{P}_{k}\,,
\end{equation}

\noindent
where $\mathcal{P}_{k}$ satisfies the equation

\begin{equation}
 d \mathcal{P}_{k}=-\imath_{k} \mathcal{F}\,.
\end{equation}

Since in our gauge $\pounds_{k} \mathcal{A}=0$ it is clear that a solution (as
a matter of fact, any solution) of the above equation is provided by

\begin{equation}
  \label{mmA}
\mathcal{P}_{k}=\imath_{k}\mathcal{A}+C\,,
\end{equation}

\noindent
for some constant $C$. Notice, though, that this is not the definition of the
momentum map, but rather a particular form of $\mathcal{P}_{k}$ which is
available in the gauge in which the black-ring solution is given. The momentum
map is, by definition, gauge invariant. The constant $C$ is determined by
demanding $\mathcal{P}_{k}$ (which will be interpreted as the black ring's
electrostatic potential $\Phi$) to vanish at infinity, and it is not difficult
to see that $C=0$.

This solution admits an analytic prolongation to the bifurcate ring
$\mathcal{BH}$ at $y=-1/\nu$ (and actually beyond that) and, in agreement with
the generalised zeroth law, it is a constant over the whole event horizon
$\mathcal{H}$ that we will denote by $ \Phi_{\mathcal{H}}$

\begin{equation}
  \begin{aligned}
    \mathcal{P}_{k}
    & \stackrel{\mathcal{H}}{=}
    \mathcal{P}_{k}(x,-1/\nu)
    \\
    & \\
    & =
    -\frac{ \cosh{2 \alpha} \left[C_{\lambda} (\mu +\nu )+3 C_{\mu}
        (\lambda -\nu )\right]+C_{\lambda} (\mu +\nu )+C_{\mu} (\lambda -\nu
      )}{\cosh{2 \alpha} \left[C_{\lambda} (\mu +\nu )+3 C_{\mu} (\lambda
        -\nu )\right]+C_{\lambda} (\mu +\nu )-3 C_{\mu} (\lambda -\nu )}\tanh{\alpha}\\
    & \\
    & \equiv
    \Phi_{\mathcal{H}}\,. 
  \end{aligned}
\end{equation}

Observe that, in the gauge in which the solution is given, the potential
$\mathcal{A}$ is ill-defined over $\mathcal{BH}$: $\imath_{k}\mathcal{A}$ is a
non-vanishing constant there and $k$ vanishes, which implies that
$\mathcal{A}$ must diverge there. It is worth stressing that the momentum map
is unaffected by such gauge pathologies since the solution Eq.~(\ref{mmA})
extends from infinity all the way down to $\mathcal{BH}$ (and beyond). This is
a consequence of the fact that, although the momentum maps may only exist
locally, they are defined by a gauge invariant equation.

The KR momentum map 1-form, $P_{k}$, is defined by Eq.~(\ref{eq:KRPkdef}),
and, for this particular solution

\begin{equation}
  \label{KReq}
  d P_{k}
   =
    -\left(\imath_{k}H+\mathcal{P}_{k}{}^{I}\mathcal{F}_{I}\right)
    =
    -\left(\imath_{k}\star \mathcal{F} +2\mathcal{P}_{k} \mathcal{F} \right)\,.
\end{equation}

If we knew the KR potential $B$ in a gauge in which $\pounds_{k}B=0$, using
$\mathcal{P}_{k}=\imath_{k}\mathcal{A}$, we would obtain the KR momentum map
1-form 

\begin{equation}\label{form1}
P_{k}=\imath_{k}B-\mathcal{P}_{k}\mathcal{A}+\alpha\,,
\end{equation}

\noindent
where $\alpha$ is an arbitrary closed 1-form, $d \alpha=0$, that could be
determined by imposing regularity: smoothness of $P_{k}$ both at the axis of
the ring, $P_{\psi}(x,y=-1)=0$, and at the outer axis of the spheres,
$P_{\varphi}(x=-1,y)=0$, so that it is well defined when approaching
infinity). Finding $B$ is, however, as hard as finding $P_{k}$ directly from
Eq.~(\ref{KReq}), which is what we are going to do, taking into account that
we are only interested in the pullback of $P_{k}$ to the constant-$t$ surface
$\Sigma$, which must be of the form

\begin{equation}
  \label{form2}
P_{k}\stackrel{\Sigma}{=} P^{\Sigma}_{k\,\varphi}(x,y) d \varphi+P^{\Sigma}_{k\,\psi}(x,y) d \psi\,,
\end{equation}

\noindent
because of the general form of the solution.

The two functions $P^{\Sigma}_{k\,\varphi}(x,y)$ and $P^{\Sigma}_{k\,\psi}(x,y)$ are given by

\begin{subequations}
  \begin{align}
    P^{\Sigma}_{k\,\varphi}(x,y)
    &=
      -\int^{y}\left(\imath_{k}\star \mathcal{F} +2\mathcal{P}_{k} \mathcal{F}
      \right)_{y\varphi}dy+f_{\varphi}(x)
      \nonumber \\
    & \nonumber \\
    &
      =-2\mathcal{P}_{k}\mathcal{A}_{\varphi}+\int^{y}I_{\varphi}(x,y)dy+f_{\varphi}(x)\, ,
    \\
    & \nonumber \\
    P^{\Sigma}_{k\,\psi}(x,y)
    &=
      -\int^{y}\left(\imath_{k}\star \mathcal{F} +2\mathcal{P}_{k} \mathcal{F}
      \right)_{y\psi}dy
      +f_{\psi}(x)
      \nonumber \\
    & \nonumber \\
    & =-2\mathcal{P}_{k}\mathcal{A}_{\psi}+\int^{y}I_{\psi}(x,y)dy+f_{\psi}(x)\, , 
  \end{align}
\end{subequations}

\noindent
where 

\begin{subequations}
  \begin{align}
    I_{\varphi}(x,y)
    & =
      2 \mathcal{A}_{\varphi} \left(\partial_{y}\mathcal{A}_{t}+\Omega
      \partial_{y}\mathcal{A}_{\psi}\right)
      \nonumber \\
    & \nonumber \\
    & \hspace{.5cm}
      +\partial_{x}\mathcal{A}_{t} \left(\frac{R^{2} \Omega  F(x) G(x) H(y)
      h(x,y)^{2}}{F(y) H(x) (x-y)^{2}}
      +\frac{F(y) G(x) H(y) \omega _{\psi} (y) (\Omega  \omega_{\psi}
      (y)+1)}{F(x) G(y) H(x) h(x,y)}\right)
      \nonumber \\
    & \nonumber \\
    & \hspace{.5cm}
      -\partial_{x}\mathcal{A}_{t} \left(\frac{\Omega  H(x)^{2}
      \omega_{\varphi} (x)^{2}}{H(y)^{2} h(x,y)}\right)
      \nonumber \\
    & \nonumber \\
    & \hspace{.5cm}
      -\partial_{x}\mathcal{A}_{\psi}\frac{F(y) G(x) H(y) (\Omega\omega_{\psi}(y)+1)}{F(x) G(y) H(x) h(x,y)}+\partial_{x}\mathcal{A}_{\varphi}\frac{\Omega  H(x)^{2} \omega_{\varphi} (x)}{H(y)^{2} h(x,y)}\,,
    \\
    & \nonumber \\
    I_{\psi}(x,y)
    & =\frac{H(x)^{2} \left(\omega_{ \varphi} (x)
      \partial_{x}\mathcal{A}_{t}-\partial_{x}\mathcal{A}_{\varphi}\right)}{H(y)^{2}
      h(x,y)}
      +2 \mathcal{A}_{\psi}\left(\partial_{y}\mathcal{A}_{t}+\Omega \partial_{y}\mathcal{A}_{\psi}\right)\,,
  \end{align}
\end{subequations}

\noindent
for some functions $f_{\varphi}(x)$ and $f_{\psi}(x)$ to be determined.

In this form, the functions are well defined at $y=-1/\nu$ (and beyond), and we
can analytically prolongate $P_{k}$ there.

The functions $f_{\varphi}(x)$ and $f_{\psi}(x)$ can be readily fixed from the
fact that the combination $P_{k}+2\mathcal{P}_{k}\mathcal{A}$ is closed on
$\mathcal{BH}$ (the restricted generalized zeroth law). Indeed, pulling back
on $\mathcal{BH}$ the KR momentum map Eq.~(\ref{KReq}), one has

\begin{equation}
  d \left(P_{k}+2\Phi_{\mathcal{H}}\mathcal{A}\right)\stackrel{\mathcal{BH}}{=}0\,.
\end{equation}

\noindent
Thus, a solution of the form \eqref{form2} that is well defined at $y=-1/\nu$
must satisfy the boundary condition

\begin{equation}
P_{k}\stackrel{\mathcal{BH}}{=}-2\Phi_{\mathcal{H}}\mathcal{A}+C_{\varphi} d \varphi+C_{\psi} d \psi
\end{equation}

\noindent
for some constants $C_{\varphi}$ and $C_{\psi}$. This implies that our
solution reads

\begin{subequations}
  \label{finalform}
  \begin{align}
    P^{\Sigma}_{k\,\varphi}(x,y)
    &=
      -2\mathcal{P}_{k}\mathcal{A}_{\varphi}+\int^{y}_{-1/\nu}I_{\varphi}(x,y)dy+C_{\varphi}\,,
    \\
    & \nonumber \\
    P^{\Sigma}_{\psi}(x,y)
    &=
      -2\mathcal{P}_{k}\mathcal{A}_{\psi}+\int^{y}_{-1/\nu}I_{\psi}(x,y)dy+C_{\psi}\,.
  \end{align}
\end{subequations}

Remarkably,

\begin{subequations}
  \begin{align}
    \int^{y}_{-1/\nu}I_{\varphi}(-1,y)dy
    & =
      0\,, \hspace{1.5cm} \forall y\neq-1\,,
    \\
    & \nonumber \\
    \int^{-1}_{-1/\nu}I_{\psi}(x,y)dy
    & = \frac{\cosh{2\alpha} \left[C_{\lambda} (\mu +\nu )+C_{\mu} (\nu
      -\lambda )\right]+C_{\lambda} (\mu +\nu )+C_{\mu} (\lambda -\nu
      )}{\cosh{2\alpha} \left[C_{\lambda} (\mu +\nu )+3 C_{\mu} (\lambda -\nu
      )\right]+C_{\lambda} (\mu +\nu )-3 C_{\mu}(\lambda -\nu )}\times
      \nonumber \\
    & \nonumber \\
    & \hspace{.5cm}
      \times\frac{\nu-1}{\mu +\nu } C_{\mu} R\ \text{sech}{\,\alpha}\,, \hspace{1.5cm}\forall x\,,
  \end{align}
\end{subequations}

\noindent
so regularity at $y=-1$ and $x=-1$ is achieved by setting

\begin{align}
  C_{\varphi}
  & = 0\,,
  \\
  & \nonumber \\
  C_{\psi}
  & =
    \frac{\cosh{2\alpha} \left[C_{\lambda} (\mu +\nu )+C_{\mu} (\nu -\lambda
    )\right]+C_{\lambda} (\mu +\nu )+C_{\mu} (\lambda -\nu )}{\cosh{2\alpha}
    \left[C_{\lambda} (\mu +\nu )+3 C_{\mu} (\lambda -\nu )\right]+C_{\lambda}
    (\mu +\nu )-3 C_{\mu}(\lambda -\nu )}\frac{1-\nu}{\mu +\nu }C_{\mu} R\
    \text{sech}{\,\alpha}
    \nonumber   \\
  & \nonumber \\
  &\equiv
    C(\lambda,\mu,\nu,\alpha)\frac{1-\nu}{\mu +\nu }C_{\mu} R\  \text{sech}{\,\alpha},
\end{align}

\noindent
which completes the solution.

We conclude by noticing that the associated KR potential 1-form at
$\mathcal{BH}$ is \textit{purely harmonic} and given by,

\begin{equation}
  \Phi_{KR}
  =
  P_{k}+2\mathcal{P}_{k}\mathcal{A}
  \stackrel{\mathcal{BH}}{=}
  \Phi_{KR\,\tilde{\psi}} d \tilde{\psi}\,,
\end{equation}

\noindent
where $\tilde{\psi}=(2\pi/\Delta\psi)\psi$ is the angular coordinate with
canonical period $\tilde{\psi}\sim\tilde{\psi}+2\pi$ and

\begin{equation}
  \Phi_{KR\,\tilde{\psi}}
  =
  C_{\psi}\frac{\Delta \psi}{2\pi}
  =
  C(\lambda,\mu,\nu,\alpha) \frac{\sqrt{1-\lambda}(1+\mu)^{3/2}}{\mu+\nu}C_{\mu} R \ \text{sech}{\,\alpha}\,.
\end{equation}

\noindent
For $\alpha=0$, $\Phi_{KR}$ coincides with the potential given in
Ref.~\cite{Emparan:2004wy} up to (parameter-independent) numerical prefactors.

\section{Discussion}
\label{sec-discussion}

In this paper we have derived the first law of black hole mechanics in the
context of the effective action of the Heterotic Superstring compactified on a
torus at leading order in $\alpha'$. The first law includes the variations of
the conserved charges of the 1-forms, $\mathcal{Q}_{I}$, and of the charges
associated to the KR field, $\mathcal{Q}_{i}$, multiplied by the potentials
$\Phi^{I}$ and $\Phi^{i}$ which are constants that we have computed on the
bifurcation surface.\footnote{It is not hard to prove that the potentials
  $\Phi^{I}$, defined as the momentum maps $\mathcal{Q}_{k}{}^{I}$ are
  constant over the complete event horizon using the dominant energy condition
  and the Einstein equations as it is done for a single 1-form field in
  Ref.~\cite{Frolov:1998wf}. It is not clear, though, how definition of the
  potentials $\Phi^{i}$ may be extended using other sections of the event
  horizon different from the bifurcation sphere because the closedness of
  $P_{k}+\mathcal{P}_{k\, I}\mathcal{F}^{I}$ is based on the property
  $\imath_{k}H\stackrel{\mathcal{BH}}{=}0$. It is not clear how to extend this
  property to other sections of the event horizon different from the
  bifurcation surface $\mathcal{BH}$.}

The main ingredients in this proof are the identification of the parameters of
the gauge transformations that generate symmetries of the complete field
configurations, the careful definitions of the associated charges and the
corresponding potentials through what we have called restricted generalized
zeroth laws.  Due to the interactions between 1-forms and the KR 2-form
induced by the Chern-Simons terms, all the terms involving charges and
potentials in the first law are interrelated and all their definitions are
either right or wrong simultaneously. This can be seen as a test of our
definitions and of the final result.

In the theory considered in this paper we have arrived at the well-known
result that the entropy is one quarter of the area. In theories of higher
order in the curvature it is known that there are additional contributions
from the terms that contain the curvature, as the Iyer-Wald prescription makes
manifest. However, as explained in the introduction, in the case of the
Heterotic Superstring effective action at first order in $\alpha'$, we also
expect that the need to have well-defined charges and, simultaneously, closed
forms over the bifurcation sphere will result in the need to include
additional terms in the ``gravitational charge'' that, in the end, will give
us the entropy. Work in this direction is well under way \cite{Elgood:2020nls}.

Finally, we would like to comment upon two apparent shortcomings of Wald's
formalism: it is not clear how to include the variation of the scalar charges
and the moduli \cite{Gibbons:1996af,Astefanesei:2018vga} in the first law. In
5 dimensions, for instance, the KR field is dual to a 1-form and black-hole
solutions electrically charged with respect to this dual 1-form exist. If we
describe the theory in terms of the KR 2-form, it is not clear how to make the
variation of this electric charge appear in the first law following this
procedure. In this particular case, the electric charge of the 1-forms would
be associated to S5-branes wrapped on T$^{5}$ and it would be very interesting
to see the precise definition of this kind of charge to try to solve the
ambiguities detected in Ref.~\cite{Faedo:2019xii}.

\section*{Acknowledgments}

TO would like to thank G.~Barnich, P.~Cano, P.~Meessen, P.F.~Ram\'{\i}rez,
A.~Ruip\'erez and C.~Shahbazi for many useful conversations and their long-term
collaboration in this research topic. DP would also like to thank G.~Barnich
for many useful conversations. This work has been supported in part by the
MCIU, AEI, FEDER (UE) grant PGC2018-095205-B-I00 and by the Spanish Research
Agency (Agencia Estatal de Investigaci\'on) through the grant IFT Centro de
Excelencia Severo Ochoa SEV-2016-0597. The work of ZE has also received
funding from ``la Caixa'' Foundation (ID 100010434), under the agreement
LCF/BQ/DI18/11660042.  The work of DP is supported by a ``Campus de Excelencia
Internacional UAM/CSIC'' FPI pre-doctoral grant.  TO wishes to thank
M.M.~Fern\'andez for her permanent support.

\appendix

\section{A truncation of the $d=5$ theory to a $\mathcal{N}=1,d=5$
  supergravity}
\label{sectruncationtoN1d5SUGRA}

A very useful, almost algorithmic, procedure has been developed in
Refs.~\cite{Gauntlett:2002nw,Gauntlett:2004qy,Gutowski:2005id,Bellorin:2007yp,Bellorin:2008we}
to construct supersymmetric solutions (black holes and black rings, in
particular) of $\mathcal{N}=1,d=5$ supergravity coupled to vector
supermultiplets.\footnote{These are supergravities invariant under 8
  independent supersymmetry transformations, which are combined in a minimal
  5-dimensional spinor. Often, they are referred to as $\mathcal{N}=2,d=5$
  supergravities.}  We can use this procedure in the context of the Heterotic
Superstring Effective action compactified on a T$^{5}$ if we find a consistent
truncation that produces a model $\mathcal{N}=1,d=5$ supergravity. A very
simple truncation with this property has been used, for instance, in
Ref.~\cite{Cano:2018qev}. It can be described more conveniently as a trivial
dimensional reduction on a T$^{4}$ (with all the fields that arise in the
reduction set to their vacuum values) followed by a non-trivial
compactification on a circle. The only fields that survive are the KR 2-form
(which can be dualized into a vector field), the KK and winding vectors and
the dilaton and KK scalars. This field content fits into $\mathcal{N}=1,d=5$
supergravity (metric and graviphoton vector field) coupled to two vector
multiplets (one vector and one real scalar field each).

In order to profit from the solution-generating techniques developed for
$\mathcal{N}=1,d=5$ supergravity theories, we need to rewrite this truncated
version of the Heterotic Superstring effective action in the appropriate form:
first, we rewrite the action in the Einstein frame and then we will dualize
the KR field into a vector. After that, we will identify the scalar manifold
etc.

The action of the truncated theory is

\begin{equation}
\label{eq:heteroticd5}
\begin{aligned}
  S[e^{a},B,\phi,k,A,B]
  & =
  \frac{g_{s}^{(5)\, 2}}{16\pi G_{N}^{(5)}} \int e^{-2\phi}
  \left[\star (e^{a}\wedge e^{b}) \wedge R_{ab}
    -4d\phi\wedge \star d\phi
  \right.
  \\
  & \\
  & \hspace{.5cm}
  \left.
      +\tfrac{1}{2}k^{-2}dk \wedge \star dk
    -\tfrac{1}{2}k^{2}F\wedge \star F
    -\tfrac{1}{2}k^{-2}G\wedge \star G
    +\tfrac{1}{2}H\wedge \star H
  \right]\, ,
\end{aligned}
\end{equation}

\noindent
where $H$ is simply

\begin{equation}
H = dB -\tfrac{1}{2}A\wedge G -\tfrac{1}{2}B\wedge F\, .  
\end{equation}

The string-frame Vielbein $e^{a}$ is related to the (modified) Einstein-frame
Vielbein $\tilde{e}^{a}$ by

\begin{equation}
  e^{a} = e^{2(\phi-\phi_{\infty})/3}\tilde{e}^{a}\, ,
  \hspace{1cm}
  g_{s} = e^{\phi_{\infty}}\, ,
\end{equation}

\noindent
and the action in the (modified) Einstein frame takes the form (removing the
tildes for simplicity)

\begin{equation}
\label{eq:heteroticd5Einsteinframe}
\begin{aligned}
  S[e^{a},B,\phi,k,A,B]
  & =
  \frac{1}{16\pi G_{N}^{(5)}} \int 
  \left[\star (e^{a}\wedge e^{b}) \wedge R_{ab}
    +\tfrac{4}{3}d\phi\wedge \star d\phi
      +\tfrac{1}{2}k^{-2}dk \wedge \star dk
  \right.
  \\
  & \\
  & \hspace{.5cm}
  \left.
    -\tfrac{1}{2}k^{2}e^{-4\phi/3}F\wedge \star F
    -\tfrac{1}{2}k^{-2}e^{-4\phi/3}G\wedge \star G
    +\tfrac{1}{2}e^{-8\phi/3}H\wedge \star H
  \right]\, .
\end{aligned}
\end{equation}

The next step is the dualization of the KR 2-form. As usual, we consider the
above action as a functional of the 3-form field strength $H$ and add a
Lagrange-multiplier term to enforce its Bianchi identity
$dH= -\tfrac{1}{2}\mathcal{F}_{I}\wedge \mathcal{F}^{I}$

\begin{equation}
\label{eq:heteroticd5Einsteinframe2}
\begin{aligned}
  S[e^{a},H,\phi,k,A,B]
  & =
  \frac{1}{16\pi G_{N}^{(5)}} \int 
  \left[\star (e^{a}\wedge e^{b}) \wedge R_{ab}
    +\tfrac{4}{3}d\phi\wedge \star d\phi
      +\tfrac{1}{2}k^{-2}dk \wedge \star dk
  \right.
  \\
  & \\
  & \hspace{.5cm}
    -\tfrac{1}{2}k^{2}e^{-4\phi/3}F\wedge \star F
    -\tfrac{1}{2}k^{-2}e^{-4\phi/3}G\wedge \star G
    +\tfrac{1}{2}e^{-8\phi/3}H\wedge \star H
  \\
  & \\
  & \hspace{.5cm}
  \left.
    -C\wedge
    \left(dH+F\wedge G \right)
  \right]\, ,
\end{aligned}
\end{equation}

\noindent
where $C$ is the 1-form dual to the 2-form $B$. Varying this action with
respect to $H$, we get

\begin{equation}
\frac{\delta S}{\delta H} = e^{-8\phi/3}\star H -dC =0\, ,  
\end{equation}

\noindent
which is solved by

\begin{equation}
H = e^{8\phi/3}\star K\, ,
  \hspace{1cm}
  K\equiv dC\, .
\end{equation}

Substituting this solution into the action
Eq.~(\ref{eq:heteroticd5Einsteinframe2}) we find the dual action

\begin{equation}
\label{eq:heteroticd5Einsteinframedual}
\begin{aligned}
  S[e^{a},\phi,k,A,B,C]
  & =
  \frac{1}{16\pi G_{N}^{(5)}} \int 
  \left[\star (e^{a}\wedge e^{b}) \wedge R_{ab}
    +\tfrac{4}{3}d\phi\wedge \star d\phi
      +\tfrac{1}{2}k^{-2}dk \wedge \star dk
  \right.
  \\
  & \\
  & \hspace{.5cm}
    -\tfrac{1}{2}k^{2}e^{-4\phi/3}F\wedge \star F
    -\tfrac{1}{2}k^{-2}e^{-4\phi/3}G\wedge \star G
    -\tfrac{1}{2}e^{8\phi/3}K\wedge \star K
  \\
  & \\
  & \hspace{.5cm}
  \left.
        -F\wedge G\wedge C
  \right]\, .
\end{aligned}
\end{equation}

The final step consists in finding the relation between the fields of
this action and those of a $\mathcal{N}=1,d=5$ theory with two vector
supermultiplets written in the standard form\footnote{Here we are
  using the notation and conventions of
  Ref.~\cite{Bergshoeff:2004kh} with minor changes explained in
  Appendix~A of Ref.~\cite{Bellorin:2006yr}. See also
  Ref.~\cite{Ortin:2015hya}.}

\begin{equation}
\label{eq:N1d5action}
\begin{aligned}
  S[e^{a},\phi^{x},A^{I}]
  & =
  \frac{1}{16\pi G_{N}^{(5)}} \int 
  \left[\star (e^{a}\wedge e^{b}) \wedge R_{ab}
      +\tfrac{1}{2}g_{xy}d\phi^{x} \wedge \star d\phi^{y}
    -\tfrac{1}{2}a_{IJ}F^{I}\wedge \star F^{J}
  \right.
  \\
  & \\
  & \hspace{.5cm}
  \left.
        +\tfrac{1}{3^{3/2}}C_{IJK} F^{I}\wedge F^{J} \wedge A^{K}
  \right]\, ,
\end{aligned}
\end{equation}

\noindent
where the indices $I,J,\ldots = 0,1,2$ and the indices $x,y,\ldots = 1,2$. The
metrics $g_{xy}(\phi),a_{IJ}(\phi)$ are defined in terms of the symmetric,
constant tensor $C_{IJK}$ which fully characterizes the theory and the
\textit{real special geometry} of the scalar manifold as follows: we start by
defining 3 combinations of the 2 scalars $h^{I}(\phi)$ that satisfy the
constraint

\begin{equation}
C_{IJK}h^{I}(\phi)h^{J}(\phi)h^{K}(\phi)= 1\, .
\end{equation}

Next, we define

\begin{equation}
\label{eq:h_I}
h_{I}\equiv C_{IJK}h^{J}h^{K},
\,\,\,\,\,
\Rightarrow
\,\,\,\,\,
h^{I}h_{I}=1,
\end{equation}

\noindent
and 

\begin{equation}
h^{I}_{x} 
\equiv
-\sqrt{3} h^{I}{}_{,x}
\equiv  
-\sqrt{3} \frac{\partial h^{I}}{\partial\phi^{x}},  
\hspace{1cm}
h_{Ix}
\equiv  
+\sqrt{3}h_{I, x},
\,\,\,\,\,
\Rightarrow
\,\,\,\,\,
h_{I}h^{I}_{x}
=
h^{I}h_{Ix}
=
0.   
\end{equation}

\noindent
Then, $a_{IJ}$ is defined implicitly by the relations

\begin{equation}
h_{I}  = a_{IJ}h^{I},
\hspace{1cm}
h_{Ix}  = a_{IJ}h^{J}{}_{x}.
\end{equation}

\noindent
It can be checked that

\begin{equation}
a_{IJ}
=
-2C_{IJK}h^{K} +3h_{I}h_{J}.  
\end{equation}

The metric of the scalar manifold $g_{xy}(\phi)$, which we will use to raise
and lower $x,y$ indices is (proportional to) the pullback of $a_{IJ}$

\begin{equation}
g_{xy}
\equiv
a_{IJ}h^{I}{}_{x}h^{J}{}_{y}
=
-2C_{IJK}h_{x}^{I}h_{y}^{J}h^{K}.
\end{equation}

If we make the identifications

\begin{equation}
  A^{0} = -\sqrt{3}C\, ,
  \hspace{1cm}
  A^{1} = -\sqrt{3}A\, ,
  \hspace{1cm}
  A^{2} = -\sqrt{3}B\, ,
\end{equation}

\noindent
we find that 

\begin{equation}
  C_{012} = 1/6\, ,
  \hspace{.5cm}
  a_{00} = e^{8\phi/3}/3\, ,
  \hspace{.5cm}
  a_{11} = k^{2}e^{-4\phi/3}/3\, ,
  \hspace{.5cm}
  a_{22} = k^{-2}e^{-4\phi/3}/3\, .
\end{equation}

Since, for this $C_{IJK}$, the only non-vanishing components of $a_{IJ}$ are
the diagonal ones with $a_{II} = 3 (h_{I})^{2}$ we find that

\begin{equation}
  h_{0} = e^{4\phi/3}/3\, ,
  \hspace{1cm}
  h_{1} = ke^{-2\phi/3}/3\, ,
  \hspace{1cm}
  h_{2} = k^{-1}e^{-2\phi/3}/3\, ,
\end{equation}

\noindent
which, in its turn, implies that

\begin{equation}
  h^{0} = e^{-4\phi/3}\, ,
  \hspace{1cm}
  h^{1} = k^{-1}e^{2\phi/3}\, ,
  \hspace{1cm}
  h^{2} = ke^{2\phi/3}\, .
\end{equation}

\noindent
Finally, the non-vanishing components of the scalar metric are

\begin{equation}
  g_{\phi\phi} = 8/3\, ,
  \hspace{1cm}
  g_{kk} = k^{-2}\, .
\end{equation}

The equations of motion of a general $\mathcal{N}=1,d=5$ theory are (up to a
global factor of $(16\pi G_{N}^{(5)})^{-1}$ that we omit for simplicity)

\begin{subequations}
  \begin{align}
  \mathbf{E}_{a}
  & =
    \imath_{a}\star (e^{c}\wedge e^{d})\wedge R_{cd}
      -\tfrac{1}{2}g_{xy}
      \left(\imath_{a}d\phi^{x} \star d\phi^{y}
      +d\phi^{x}\wedge \imath_{a}\star d\phi^{y}\right)
    \nonumber \\
    & \nonumber \\
    &
       \hspace{.5cm}
      +\tfrac{1}{2}a_{IJ}\left(\imath_{a}F^{I}\wedge \star F^{J}
      -F^{I}\wedge \imath_{a}\star F^{J}
      \right)\, ,
      \\
    & \nonumber \\
    \mathbf{E}_{x}
    & =
-g_{xy}\left\{d\star d\phi^{y}+\Gamma_{zw}{}^{y}d\phi^{z}\wedge \star
      d\phi^{w} +\tfrac{1}{2}\partial^{y}a_{IJ}F^{I}\wedge\star F^{J}
      \right\}\, ,
          \\
    & \nonumber \\
    \mathbf{E}_{I}
    & =
            -d\left(a_{IJ}\star F^{J} \right)
      +\tfrac{1}{\sqrt{3}}C_{IJK}F^{J}\wedge F^{K}\, .
 \end{align}
 \end{subequations}

In this action, $\phi$ stands, actually, for $\phi-\phi_{\infty}$. In other
words: the field $\phi$ is constrained to vanish at infinity.
 
For the particular model that we have obtained as a truncation of the
compactified Heterotic Superstring effective action in $d=5$ dimensions, these
equations take the particular form

\begin{subequations}
  \begin{align}
  \mathbf{E}_{a}
  & =
    \imath_{a}\star (e^{c}\wedge e^{d})\wedge R_{cd}
      -\tfrac{4}{3}
      \left(\imath_{a}d\phi \star d\phi
      +d\phi\wedge \imath_{a}\star d\phi\right)
    \nonumber \\
    & \nonumber \\
    &
       \hspace{.5cm}
      -\tfrac{1}{2}k^{-2}
      \left(\imath_{a}dk \star dk
      +dk\wedge \imath_{a}\star dk\right)
      +\tfrac{1}{6}e^{8\phi/3}\left(\imath_{a}F^{0}\wedge \star F^{0}
      -F^{0}\wedge \imath_{a}\star F^{0}
      \right)
    \nonumber \\
    & \nonumber \\
    &
       \hspace{.5cm}
      +\tfrac{1}{6}e^{-4\phi/3}k^{2}\left(\imath_{a}F^{1}\wedge \star F^{1}
      -F^{1}\wedge \imath_{a}\star F^{1}
      \right)
      +\tfrac{1}{6}e^{-4\phi/3}k^{-2}\left(\imath_{a}F^{2}\wedge \star F^{2}
      -F^{2}\wedge \imath_{a}\star F^{2}
      \right)\, ,
      \\
    & \nonumber \\
    \mathbf{E}_{\phi}
    & =
-\tfrac{8}{3}\left\{d\star d\phi
      +\tfrac{1}{6}e^{8\phi/3}F^{0}\wedge\star F^{0}
      -\tfrac{1}{12}e^{-4\phi/3}k^{2}F^{1}\wedge\star F^{1}
      -\tfrac{1}{12}e^{-4\phi/3}k^{-2}F^{2}\wedge\star F^{2}
      \right\}\, ,
          \\
    & \nonumber \\
    \mathbf{E}_{k}
    & =
      -k^{-2}\left\{d\star dk-k^{-1}dk\wedge \star k
      +e^{-4\phi/3}k^{3}F^{1}\wedge\star F^{1}
      -k^{-1}e^{-4\phi/3}F^{2}\wedge\star F^{2}
      \right\}\, ,
          \\
    & \nonumber \\
    \mathbf{E}_{0}
    & =
            -\tfrac{1}{3}d\left(e^{8\phi/3}\star F^{0} \right)
      +\frac{1}{3^{3/2}}F^{1}\wedge F^{2}\, ,
          \\
    & \nonumber \\
    \mathbf{E}_{1}
    & =
            -\tfrac{1}{3}d\left(e^{-4\phi/3}k^{2}\star F^{1} \right)
      +\frac{1}{3^{3/2}}F^{0}\wedge F^{2}\, ,
          \\
    & \nonumber \\
    \mathbf{E}_{2}
    & =
            -\tfrac{1}{3}d\left(e^{-4\phi/3}k^{-2}\star F^{2} \right)
      +\frac{1}{3^{3/2}}F^{0}\wedge F^{1}\, .
\end{align}
\end{subequations}

\subsection{Further truncation to pure $\mathcal{N}=1,d=5$ supergravity}
\label{sec-truncationpureN1d5SUGRA}

We can truncate this theory further, to minimal (\textit{pure}) supergravity
as follows: if the two scalars are constant, taking into account that for
$\phi$ this constant value must be $\phi=0$, (we call $k_{\infty}$ the
constant value of $k$) their equations become the constraints

\begin{subequations}
  \begin{align}
0
    & =
      F^{0}\wedge\star F^{0}
      -\tfrac{1}{2}k_{\infty}^{2}F^{1}\wedge\star F^{1}
      -\tfrac{1}{2}k_{\infty}^{-2}F^{2}\wedge\star F^{2}\, ,
          \\
    & \nonumber \\
0
    & =
      k_{\infty}^{3}F^{1}\wedge\star F^{1}
      -k_{\infty}^{-1}F^{2}\wedge\star F^{2}\, ,
  \end{align}
\end{subequations}

\noindent
whose simplest solution is this relation between vector field strengths

\begin{equation}
F^{0}= k_{\infty}F^{1} = k_{\infty}^{-1}F^{2} \equiv F\, .  
\end{equation}

\noindent
Substituting this solution into the Einstein and vector equations we get only
these two independent equations

\begin{subequations}
  \begin{align}
  \mathbf{E}_{a}
  & =
    \imath_{a}\star (e^{c}\wedge e^{d})\wedge R_{cd}
      +\tfrac{1}{2}\left(\imath_{a}F\wedge \star F -F\wedge \imath_{a}\star F
      \right)
      \\
    & \nonumber \\
   -\tfrac{2}{3}\mathbf{E}
    & =
            -\tfrac{1}{3}d\star F +\frac{1}{3^{3/2}}F\wedge F\, ,
\end{align}
\end{subequations}

\noindent
which follow from the action of minimal $d=5$ supergravity
\cite{Cremmer:1980gs}

\begin{equation}
\label{eq:pureN1d5action}
\begin{aligned}
  S[e^{a},A]
  & =
  \frac{1}{16\pi G_{N}^{(5)}} \int 
  \left[\star (e^{a}\wedge e^{b}) \wedge R_{ab} -\tfrac{1}{2}F\wedge \star F
        +\tfrac{1}{6\sqrt{3}} F\wedge F \wedge A
  \right]\, .
\end{aligned}
\end{equation}

The truncation procedure we have followed to arrive to this action starting
from the 10-dimensional Heterotic Superstring effective action can be easily
reversed to embed solutions of pure $\mathcal{N}=1,d=5$ supergravity into the
10-dimensional Heterotic Superstring effective theory. In particular, we apply
this recipe to the charged, non-extremal, black ring solution of
Ref.~\cite{Elvang:2004xi} in Section~\ref{sec-example}.


\end{document}